# Plasmonic Photocatalysis Enables Selective Oxidative Coupling of Methane with Nitrous Oxide under Ambient Conditions

Serin Lee†[1], Lin Yuan†[1], Elijah Begin[2], Dali Yang[3], Cedric Lim[1], Yirui Arlene Zhang[1], Lu Ma[3], Colin Ophus[1], Yi Cui[1], Junwei Lucas Bao[2]*, Jennifer A. Dionne[1,4]*

1. Department of Materials Science and Engineering, Stanford University School of Engineering, Stanford, CA, USA.
2. Department of Chemistry, Boston College, Chestnut Hill, MA, USA.
3. Brookhaven National Laboratory, Upton, NY, USA.
4. Department of Radiology, Stanford University School of Medicine, Stanford, CA, USA.

**Abstract**

Methane ($CH_4$) and nitrous oxide ($N_2O$) are potent greenhouse gases that represent substantial chemical energy. Conversion of these abundant waste gases to high-value chemicals typically requires high temperatures up to 1000 °C, producing substantial $CO_2$ emissions and limited selectivity toward desirable multi-carbon products. Here we demonstrate a plasmonic photocatalyst that enables $CH_4$ and $N_2O$ conversion under ambient conditions to form $C_2$ and $C_3$ hydrocarbons. By systematically tuning AuPd alloys on $TiO_2$, we identify an optimal composition ($AuPd_{0.05}$) where Au enhances light harvesting and Pd enables selective C–H activation and C–C coupling. Under visible-light illumination, this catalyst produces $C_2H_4$, $C_2H_6$, $C_3H_6$, and $C_3H_8$ with ~80% selectivity while suppressing $CO_2$ formation. *In-situ* spectroscopy and hot-carrier calculations show that plasmon-generated carriers redistribute interfacial hydroxyl intermediates, shifting the hydrophilic center to suppress overoxidation. Ab-initio calculations further reveal the reduction in C-C coupling barriers from 2.7 eV to 0.7 eV under illumination. Our work illustrates how engineering interfacial electronic and adsorbate dynamics enables selective multicarbon formation.

**Introduction**

Methane ($CH_4$) and nitrous oxide ($N_2O$) are the most potent greenhouse gases, with global warming potentials 28 and 300 times that of carbon dioxide ($CO_2$), respectively, and atmospheric lifetimes ranging from decades to over a century. [1,2] Despite their environmental burden, both molecules are chemically energy-rich: $CH_4$ is an abundant carbon feedstock, whereas $N_2O$ is an oxygen-rich oxidant. Their coupled conversion into value-added products under mild conditions is therefore highly attractive, but remains a long-standing challenge in catalysis due to the intrinsic stability of $CH_4$ and the difficulty of controlling oxidation pathways. The strong C–H bond (439 kJ mol$^{-1}$) and symmetric tetrahedral geometry of $CH_4$ hinder activation under mild conditions.[3,4] Industrial $CH_4$ conversion therefore relies on energy-intensive processes such as steam methane reforming (700–1000 °C, 3–35 bar) and Fischer–Tropsch synthesis (200–300 °C, 10–60 bar),[5] which contribute significantly to $CO_2$ emissions,[6,7] while even direct oxidative or non-oxidative coupling of $CH_4$ requires harsh conditions exceeding 400 °C and 10 bar.[8–10] In contrast, $N_2O$ remains largely overlooked in heterogeneous catalysis and is primarily employed as a stoichiometric oxidant in laboratory-scale oxidation reactions.[11] The activation of both molecules requires nanostructure engineering to control the local confinement and coordination at transition-metal nanoparticles, thereby enabling the coupled reaction pathways.

Photocatalysis provides access to non-equilibrium surface states at the nanoscale, enabling reaction pathways inaccessible under thermodynamic control.[12–28] Both semiconductor-based and plasmonic photocatalysts have demonstrated the ability to promote C–H activation and C–C coupling under visible-light illumination.[3,4,29] These light-driven processes create opportunities to steer methane conversion through either non-oxidative coupling of methane (NOCM)[10,30–35] or oxidative coupling of methane (OCM) [36–44]. However, both routes retain fundamental limitations. NOCM avoids overoxidation but remains thermodynamically unfavorable (ΔG° ≈ 68.6 kJ mol$^{-1}$) and prone to coking (ΔG° ≈ 50.7 kJ mol$^{-1}$).[45] OCM alleviates these thermodynamic barriers but typically relies on $O_2$, which generates highly reactive oxygen species that promote overoxidation to carbon monoxide/carbon dioxide.[45,46]

Recent thermocatalytic studies have shown that replacing $O_2$ with softer oxidants, such as N2O, can improve selectivity by favoring the formation of surface-bound monoatomic oxygen species while suppressing unselective gas-phase oxygen chemistry.[47] Unlike $O_2$, $N_2O$ decomposes on catalyst surfaces to generate these more selective oxygen species, thereby limiting $CO_x$ formation and enhancing $C_2$ selectivity.[48] However, $N_2O$-based thermocatalytic systems still exhibit limited overall $C_2$ yields, rarely exceeding ~30%.[47,48] More importantly, despite the advantages of $N_2O$ as a milder oxidant, its use in photocatalytic methane conversion remains largely unexplored. This gap highlights an opportunity to combine the selective oxygen chemistry enabled by $N_2O$ with the non-equilibrium activation pathways accessible under light-driven conditions.

Here, we demonstrate a plasmonic photocatalytic strategy that converts $CH_4$ and $N_2O$ into higher-value hydrocarbons under ambient conditions, achieving selective formation of $C_2$ and $C_3$ products under white-light illumination. Our AuPd bimetallic alloy supported on $TiO_2$ integrates efficient plasmonic light harvesting from Au with catalytic C–H activation and C–C coupling at Pd sites, enabling the coordinated utilization of two potent greenhouse gases within a single framework. We vary the optimized composition of AuPd/TiO$_2$ and the CH$_4$:N$_2$O flow ratio, and find that AuPd0.05/TiO$_2$ exhibits the highest selectivity of 80% for C$_2$ and C$_3$ products relative to CO or CO$_2$. Under thermal bias without light, only CO$_2$ is generated as the product via overoxidation. By correlating three-dimensional structural characterization with spectroscopic signatures of composition-dependent alloying and interfacial Pd–O coordination, we establish the structural origins of reactivity and selectivity across AuPd/TiO$_2$ catalysts. Combining *in-situ* diffuse reflectance infrared Fourier-transform spectroscopy (DRIFTS) with ground- and excited-state theoretical modeling, we show that plasmon excitation actively redistributes surface hydroxyl species at the nanoscale interface, shifting the hydrophilic center and reconfiguring the catalytic interface toward selective C–C coupling rather than overoxidation. This selectivity toward oxygen-free hydrocarbons through C-C coupling emerges exclusively under illumination and cannot be achieved under dark, thermally driven conditions. Furthermore, we identify that precise control of early-stage CH$_3$* coupling is critical in directing selectivity toward multicarbon products. Together, these findings establish a framework in which light, bimetallic alloy design, and oxidant chemistry act in concert to overcome the conventional selectivity limitations in methane conversion.

**Results and discussion**
**Characterization of the catalyst structure**
Photocatalytic conversion of methane using nitrous oxide as an oxidizing agent occurs on AuPd/TiO$_2$ catalysts, as shown in Fig. 1a. We utilize a commercial TiO$_2$ support, in a mixture of polymorphs (P25, anatase and rutile, Fisher Scientific). This material serves as both the substrate and matrix to host the Au and Pd seeds, which become small (~4 nm) nanoparticles. These AuPd$_x$ (Au:Pd molar ratio = 1:x) nanoparticles supported on TiO$_2$ (AuPd$_x$/TiO$_2$) are synthesized using a co-deposition-precipitation method. Briefly, Au$^{3+}$ and Pd$^{2+}$ precursors were impregnated onto oxygen-deficient TiO$_2$ using urea as a reducing and precipitating agent, followed by reduction of the supported metals in flowing H$_2$ at 400 °C. Additional details on the dry support, codeposition of metal precursor on oxide matrix, and the *in-situ* reduction in hydrogen is described in the Methods. Fig. 1b-1k shows the structural characterization of a representative catalyst composition AuPd$_{0.05}$/TiO$_2$, with a 3 wt% loading. The bright-field transmission electron microscopy (TEM) images of the AuPd$_{0.05}$/TiO$_2$ sample (Fig. 1b) reveal the lattice spacings of the AuPd nanoparticles and the TiO$_2$ support, which exhibit anatase and rutile phases.
Tomography reconstruction based on the HAADF-STEM image further confirms the supported catalyst structure in three dimensions, revealing the spatial distribution and interfacial contact of the nanoparticles with the support that are not apparent from projection images alone (Fig. 1c). Particle size statistics from the tomography reconstruction data show the average diameter of 4.0 nm (Supplementary Fig. 1). High-angle annular dark field scanning transmission electron microscope (HAADF-STEM) imaging (Fig. 1d) reveals that nanoparticles are homogeneously distributed over the TiO$_2$ support, and the corresponding energy dispersive X-ray (EDX) map also confirms the alloying of AuPd nanoparticles (Fig. 1e, Supplementary Fig. 2).

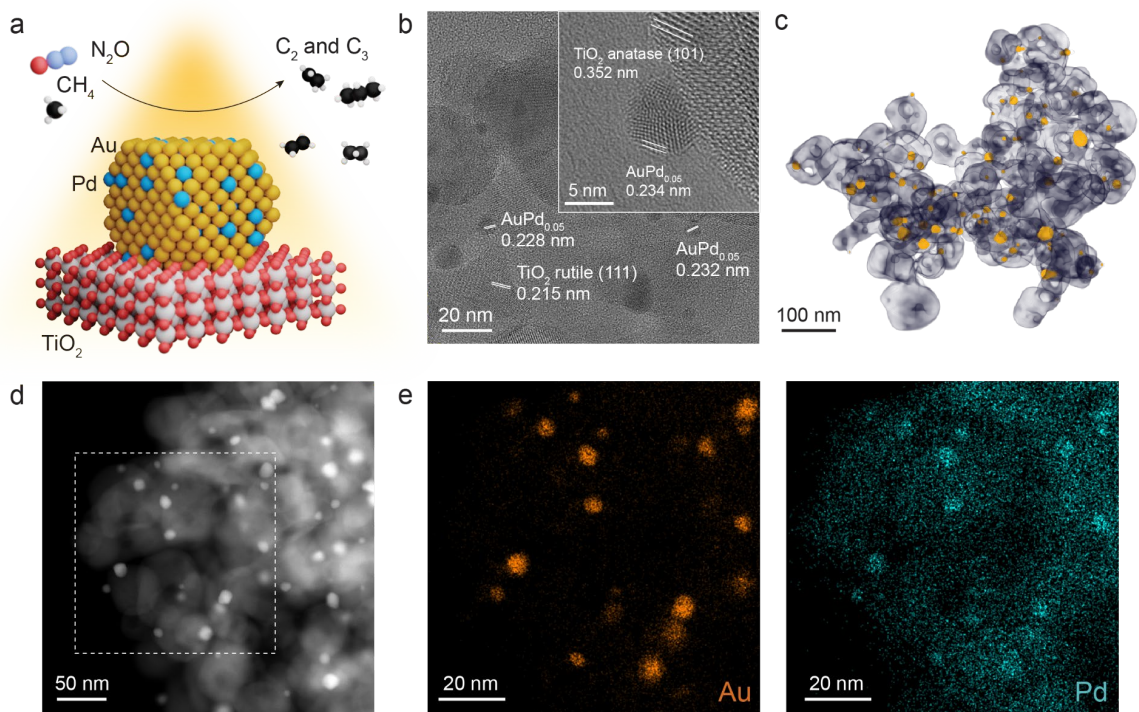

**Figure 1. Schematic of the photocatalytic methane conversion with nitrous oxide and its characterization** (a) A schematic diagram of the methane conversion over AuPd bimetallic alloy on $TiO_2$ support. (b) High-resolution TEM image of $AuPd_{0.05}/TiO_2$. The inset shows a representative magnified image of the AuPd bimetallic alloy on $TiO_2$ support. (c) Tomographic reconstruction from HAADF-STEM showing the internal structure of the $AuPd_{0.05}/TiO_2$. (d, e) HAADF-STEM image of $AuPd_{0.05}/TiO_2$ (d) and corresponding EDS elemental maps of the dashed area (e), where each map corresponds to Au and Pd, respectively.

Fig. 2a shows the diffuse reflectance UV-vis spectra of $AuPd_x/TiO_2$ catalysts with varying Pd molar ratios (0.01, 0.05, 0.1, 0.2, and 0.3) and a Pd-only control. The Pd-only sample contains 0.06 wt% Pd, matching the total metal loading of the $AuPd_{0.05}$ catalyst to enable a direct comparison of optical responses at identical metal content. The spectra exhibit a broad plasmon resonance centered at 500–600 nm, characteristic of Au nanoparticles. As the Pd fraction increases beyond $x = 0.1$, the plasmon band becomes progressively broadened and damped due to the higher dielectric losses of Pd and the small lattice mismatch that favors the formation of fcc-type Au–Pd alloys. Fig. 2b shows Monte Carlo simulations of 300 randomly sized alloy nanoparticles embedded in a $TiO_2$ matrix, modeled using Mie theory. The inset of Fig. 2b and Supplementary Fig. 3 are the near-field simulations of the electric field at the wavelength of 550 nm, visualizing the decrease in the field density as the Pd fraction increases. The simulated ensemble spectra reproduce the experimental broadening trend, confirming that the observed damping originates primarily from Au–Pd alloying. Pd K-edge XANES (Fig. 2c) shows a pronounced white-line feature and a higher edge energy for $Pd/TiO_2$ relative to Pd foil, consistent with a more oxidized and oxygen-coordinated Pd environment. Upon introducing Au, the white-line intensity progressively decreases and the edge shifts to lower energy, indicating that Pd becomes increasingly metallic within an Au–Pd alloy environment and that Pd–O coordination is reduced and/or weakened by alloying-driven electronic interaction between Au and Pd. Fourier-transformed Pd K-edge EXAFS (Fig. 2d, See Supplementary Table 1 for the details) further quantifies this structural evolution. For AuPd$_{0.01}$ and AuPd$_{0.05}$, the spectra are dominated by a Pd–Au first-shell contribution with fitted bond lengths of ~2.825–2.839 Å, consistent with Pd being embedded in an Au-rich alloy coordination. As the Pd fraction increases (AuPd$_{0.10}$–AuPd$_{0.30}$), an additional Pd–O contribution becomes measurable with fitted Pd–O distances of ~2.00–2.04 Å, indicating the emergence of oxygen-bound Pd sites at the metal–oxide interface alongside residual Pd–Au coordination. Although the fitting does not completely exclude the presence of Pd-O coordination at low Pd fraction, the strong Pd-Au contribution with no distinct Pd-O peak indicates strong alloying and a reduced affinity of Pd toward O for lower Pd fraction catalysts. In contrast, Pd/TiO$_2$ is primarily described by Pd–O coordination with a bond distance of ~2.04 Å, consistent with surface/ interfacial Pd–O species. Collectively, these results establish composition-dependent alloying and interfacial Pd–O coordination that provide a structural basis for interpreting the distinct reactivity/selectivity trends across AuPd/TiO$_2$ catalysts.

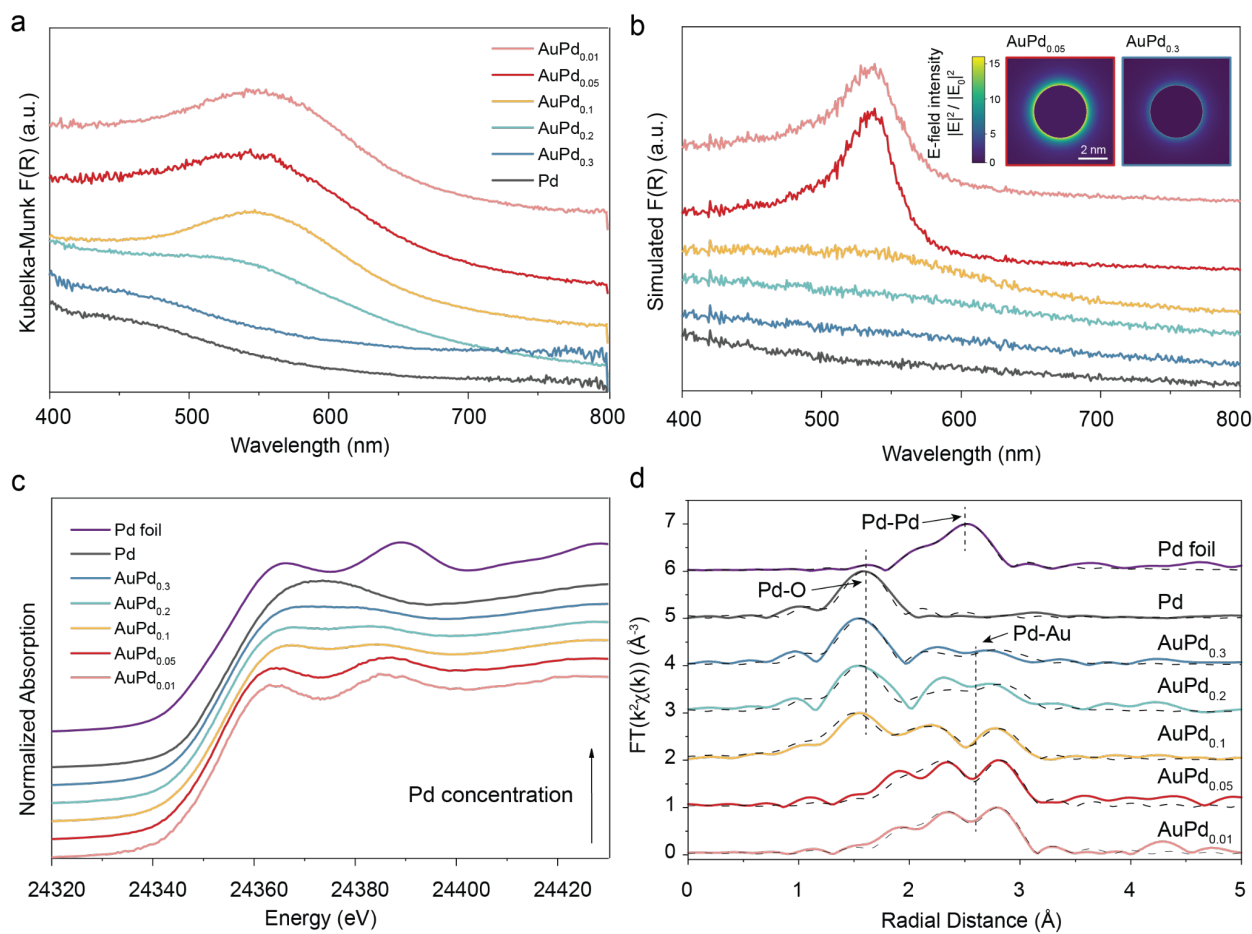

**Figure 2. Spectroscopic characterization of AuPd$_x$/TiO$_2$ catalysts.** (a,b) Normalized diffuse reflectance extinction spectra for (a) experiment and (b) Monte-Carlo simulation of the spectra in (a) on the AuPd$_x$ supported on TiO$_2$. The inset images of (b) show the near-field simulation of the electric field at wavelengths of 550 nm for AuPd$_{0.05}$ and AuPd$_{0.3}$. The dielectric function of AuPd alloy was adapted from a previously reported tool based on the first-principles calculations. (c) Pd K-edge XANES spectra showing the evolution of the near-edge structure as a function of Pd composition, indicating progressive changes in the local electronic environment of Pd atoms and alloy formation with Au. (d) Fourier-transformed EXAFS spectra (FT(k²χ(k))) revealing Pd–Au and Pd–O coordination shells. Solid lines: experimental data; Dashed lines: fitted data from bond information.

**Catalytic Measurements of Methane Conversion**

Photocatalytic methane conversion was evaluated using 15 mg of 3 wt% AuPd$_{0.05}$/TiO$_2$ under white-light illumination (400–800 nm, 500 mW cm$^{-2}$) (Fig. 3a). The catalyst was loaded into a stainless-steel Harrick cell and continuously supplied with a controlled gas mixture of CH$_4$, N$_2$O, and Ar, where flow ratios (for example, 20:2:10) correspond to 20, 2, and 10 sccm, respectively. Illumination was provided by a supercontinuum laser equipped with a tunable bandpass filter (SuperK Varia, NKT Photonics), and the surface temperature of the catalyst bed was monitored in real time by an infrared camera (FLIR A700) through a ZnSe window. Reaction products were quantified by a gas chromatography reactor (see Methods for the details). The catalytic measurements were conducted using an open-source automated photocatalytic reactor package connected to the mass flow controller, laser sources, and heating systems.[49]

Under optimized conditions (CH$_4$:N$_2$O:Ar = 20:2:10), the catalyst produced C$_2$-C$_3$ hydrocarbons, including C$_2$H$_4$, C$_2$H$_6$, C$_3$H$_6$, and C$_3$H$_8$, with high selectivity about 80%. Without Ar dilution (20:2), the selectivity declined to 60%, while increasing the N$_2$O ratio (20:5:10) caused overoxidation to CO$_2$ and CO; at CH$_4$:N$_2$O:Ar = 20:10:10, only CO$_2$ and CO were formed. We observed the surface temperature of the catalysts stayed at ~400-500 °C using infrared imaging (Supplementary Fig. 4). When heat is added to the optimized condition (CH$_4$:N$_2$O:Ar = 20:2:10) with the white-light illumination, thermal-photocatalytic reactions between 300 °C and 500 °C yielded only CO$_2$ and CO (Supplementary Fig. 5a). For catalysts with different compositions, even with the optimized condition, only overoxidation took place with AuPd$_{0.3}$/TiO$_2$ and Pd/TiO$_2$, whereas AuPd$_{0.01}$/TiO$_2$, AuPd$_{0.1}$/TiO$_2$, and AuPd$_{0.2}$/TiO$_2$ showed no detectable methane conversion (Supplementary Fig. 5b). In the absence of illumination, the thermocatalytic reaction only yielded CO$_2$ at every catalyst composition under the optimized reactant gas condition (CH$_4$:N$_2$O:Ar = 20:2:10) (Fig. 3b, Supplementary Fig. 5c).

***In-situ* Vibrational Spectroscopy and Reaction Mechanism**

We employed *in-situ* DRIFTS measurements under the same abovementioned reaction conditions for photocatalysis and thermocatalysis, to probe how light and heat drive intermediate formation (Figs. 3c–f). Under white-light illumination at CH$_4$:N$_2$O:Ar = 20:2:10, four distinct surface hydroxyl (OH*) species - terminal free (~3730 cm$^{-1}$), bridging (~3670 cm$^{-1}$), molecular (~3610 cm$^{-1}$), and adsorbed (~3440-3510 cm$^{-1}$) - were observed, where adsorbed indicates OH* with a hydrogen bond.[50,51] In contrast, at 500 °C in the dark, adsorbed OH* features were diminished (Fig.3c vs Fig. 3d). Among all compositions tested, AuPd$_{0.05}$/TiO$_2$ exhibited the strongest adsorbed OH* signal. In the 1800–1400 cm$^{-1}$ region, C–O vibrational modes together with the bending mode of $H_2O$ (~1800–1600 cm$^{-1}$) are observed in both light-driven and dark reactions at 500 °C. However, the peak intensities of these CO- and $H_2O$-associated modes are noticeably reduced under photocatalytic conditions relative to thermocatalysis (Fig. 3e and 3f). A rich catalogue $CH_x$ *, and $CH_3$ *adsorbates are observed across all compositions under photocatalytic conditions, while being most prominent for $AuPd_{0.05}/TiO_2$ (Fig. 3e). In contrast, under thermocatalytic conditions, these species are only prominently detected for the $AuPd_{0.3}$ composition (Fig. 3f), This difference suggests that, under light irradiation, more complex C–C coupling pathways are activated and compete with the direct oxidation reaction. Detailed analysis of the vibrational modes is provided in Supplementary Table 2.

Combined with the reactivity and selectivity observation in the product measurement, our results suggest that the adsorbed OH* group (hydrogen-bonded OH*) is key to regulating the selectivity. In particular, this group enables the C-C oxidative coupling process, competing with the overoxidation on the surface on $AuPd_{0.05}$ samples. As shown above, Pd K-edge EXAFS reveals composition-dependent Pd–O coordination alongside Pd–Au alloying, consistent with an interfacial Pd–O environment that could host hydroxylated/oxygenated species during reaction. The combined reactivity and selectivity trends reveal that light irradiation fundamentally reshapes the interfacial chemistry of $AuPd/TiO_2$. Rather than simply accelerating methane oxidation, illumination generates adsorbed OH* species with hydrogen bonds at the $Pd/TiO_2$ interface that act as selectivity-directing sites. These interfacial OH* groups stabilize surface-bound carbon intermediates and suppress their rapid overoxidation to $CO_2$. As a result, carbon species persist long enough to undergo C–C bond formation, enabling oxidative coupling pathways that are otherwise disfavored under purely thermal conditions. Under thermal bias, complete oxidation typically dominates because surface oxygen species rapidly convert intermediates into $CO_2$ before coupling can occur. Light-driven restructuring therefore, shifts the reaction landscape from deep oxidation toward kinetically enabled C–C coupling.

*In-situ* DRIFTS measurements further indicate that methane coupling proceeds through a Langmuir–Hinshelwood (L-H) pathway, in which two surface-adsorbed intermediates react directly with each other.[52] This assignment is supported by the simultaneous emergence and correlated evolution of adsorbed hydrocarbon ($CH_x$) and hydrogen-bonded OH vibrational features under illumination, indicating that reactive species remain stabilized on the surface prior to C–C bond formation. The persistence of these adsorbate signatures without rapid depletion suggests that surface-bound intermediates participate directly in coupling reactions rather than being immediately consumed by lattice oxygen. This contrasts with the Mars–van Krevelen (MvK) mechanism, where lattice oxygen atoms serve as reactive intermediates and are subsequently replenished from the bulk oxide, which would be expected to exhibit distinct DRIFTS fingerprints with lattice-oxygen participation such as transient depletion of Ti-O vibrational features.[53] Although Mars–van Krevelen chemistry has been reported for Pd single atoms embedded within $TiO_2$ matrices, [10,54] our spectroscopic evidence suggests that, in $AuPd/TiO_2$ under light irradiation, coupling is governed by interactions between adsorbed species at a dynamically restructured metal–oxide interface. Together, these results reveal that light does not merely enhance activity, but actively redirects the dominant reaction pathway by engineering interfacial hydrogen-bonding motifs that favor selective C–C bond formation over thermodynamically favored overoxidation.

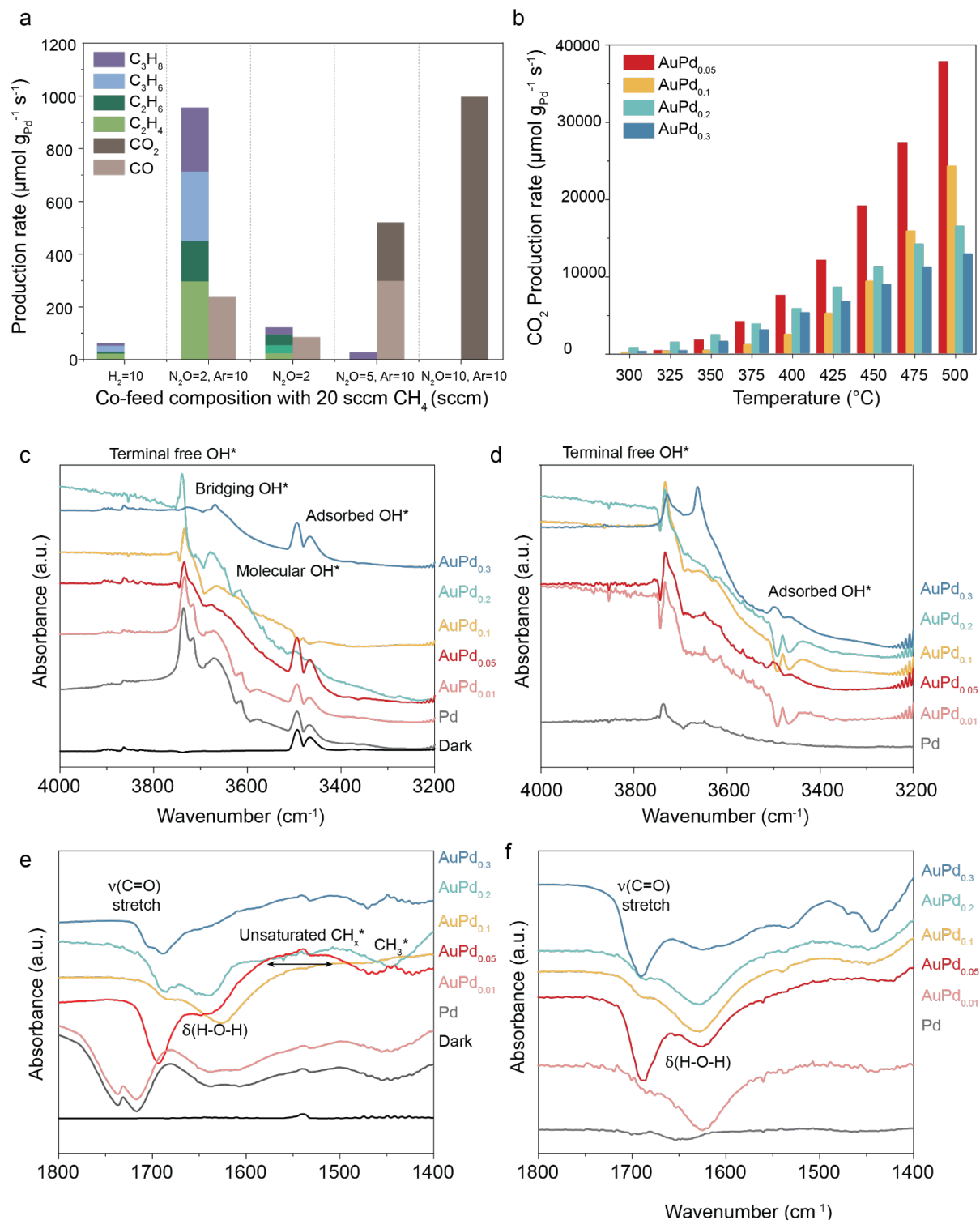

**Figure 3. Photocatalytic methane conversion performance and *in-situ* DRIFTS.** (a)

Photocatalytic methane conversion rates using 3 wt% AuPd$_{0.05}$/TiO$_2$ catalysts at different reactant gas concentrations by illuminating 400-800 nm white light at 500 mW · cm$^{-2}$. (b) Thermocatalytic methane conversion rate using 3 wt% AuPd$_x$ /TiO$_2$ (x=0.05, 0.1, 0.2, 0.3) catalysts under reactant gas concentration of CH$_4$:N$_2$O:Ar=20:2:10. (c-f) *In-situ* DRIFTS of methane conversion with the reactant gas concentration of CH$_4$:N$_2$O:Ar=20:2:10 (c,e) and thermocatalytic conditions at 500 ºC with the reactant gas concentration of CH$_4$:N$_2$O:Ar=20:2:10 (d,f)

**Wavelength-Dependent Reaction Intermediates and Plasmon-Mediated Shifting of the Hydrophilic Center**

To elucidate the role of plasmon resonance in driving the reaction mechanism, we performed wavelength-dependent *in-situ* DRIFTS measurements on AuPd$_{0.05}$/TiO$_2$. We used an illumination intensity of 100 mW · cm$^{-2}$, and a series of bandpass filters with a 50 nm bandwidth to isolate different excitation wavelengths across the visible range(490-825 nm). The max power is 80% lower than the white light illumination experiments with a narrowed wavelength range. The limited power of our laboratory laser system precluded direct detection of final reaction products (e.g., C$_2$ and C$_3$ hydrocarbons and CO/CO$_2$). Nevertheless, the vibrational features observed on AuPd$_{0.05}$/TiO$_2$ closely resembled those obtained under high-intensity white-light illumination (500 mW · cm$^{-2}$). These include the terminal-free OH* (~3700 cm$^{-1}$), adsorbed OH* (3500–3200 cm$^{-1}$), and surface-bound C=O, CH$_x$*, and CH$_3$* species (Supplementary Table. 2). Hence, the same surface reaction mechanism operates under both illumination power conditions for selective, light-driven methane coupling, which is otherwise unobservable in the dark.

The wavelength-dependent *in-situ* DRIFTS spectra acquired under illumination from 490 to 825 nm are shown in Fig. 4a, while the temperature-dependent spectra collected between 300 and 500 °C are presented in Fig. 4c. Both the spectra and the corresponding heat map (Supplementary Fig. 6) reveal distinct vibrational features at wavenumbers characteristic of terminal-free OH* and adsorbed OH* species. Previous studies have attributed the overoxidation of methane on Pd/TiO$_2$ to oxygen originating from interfacial OH* and water species, which act as oxidative agents during methane coupling.[55] To probe this process, we quantified the surface coverage of terminal-free and hydrogen-bonded adsorbed OH* by integrating their respective peak areas (Supplementary Fig. 7a,b). The area ratio of terminal-free OH* to adsorbed OH* exhibits a distinct maximum at ~550 nm (Fig. 4b), consistent with the plasmon resonance peak of AuPd, indicating the redistribution of the OH* under light illumination. In contrast, the temperature-dependent OH ratio* follows an exponential increase (Fig. 4 d, Supplementary Fig. 7c, d), mirroring the thermal reactivity trend.

To exclude possible photothermal and charge-transfer effects from TiO$_2$ to the metal, we monitored the surface temperature across wavelengths and found a mild rise in the blue–green region (490–520 nm), opposite to the OH ratio trend* (Supplementary Fig. 8). Furthermore, a control experiment using a 450 nm LED diode (600 mW · cm$^{-2}$) as the sole light source yielded no detectable CH$_x$ * or CH$_3$ * intermediates (Supplementary Fig. 9). Together, these results indicate that plasmon-mediated excitation near 550 nm governs the redistribution of OH* species, effectively shifting the hydrophilic center and steering the surface chemistry toward selective C-C coupling rather than overoxidation. We employed a semiclassical calculation of the hot-carrier cross-section by both surface-assisted Landau damping and bulk interband transition, for all $AuPd_x$ (x=0.01-0.3) compositions investigated (Calculation details in Methods and SI). We find a similar wavelength-dependent trend by comparing the normalized OH* adsorbates coverage ratio to the hot-carrier cross-section (Fig. 4e).

We hypothesize that plasmon-generated hot electrons transfer across the AuPd-$TiO_2$ interface and transiently modify the electronic structure of interfacial adsorbates through a desorption-induced-by-electronic-transition (DIET) mechanism. (Fig. 4f) [56] Such electronic excitation weakens the binding of terminal free OH* species and promotes the redistribution of surface hydroxyls toward adsorbed OH* configurations. This trend is consistent with our analysis of the terminal OH* to adsorbed OH* ratio. Under thermocatalytic conditions, the ratio remains consistently higher across the entire measured temperature range (approximately 1–4), whereas under photocatalytic conditions it is lower, varying between 0 and 1 over the examined wavelength range. Because terminal free OH* is associated with stronger oxidative reactivity compared to the adsorbed OH* forming a network of hydrogen bonds,[57] this interfacial redistribution suppresses deep oxidation to C=O products and shifts the reaction balance toward C–C coupling. This mechanism parallels prior reports on plasmon-assisted selective hydrogenation of acetylene to ethylene,[16] where hot-electron excitation stabilizes transient anionic intermediates and alters adsorbate binding energetics. In both systems, nonequilibrium carrier populations reshape the surface reaction landscape, selectively suppressing over-oxidation (or over-hydrogenation) pathways while promoting the desired coupling chemistry.

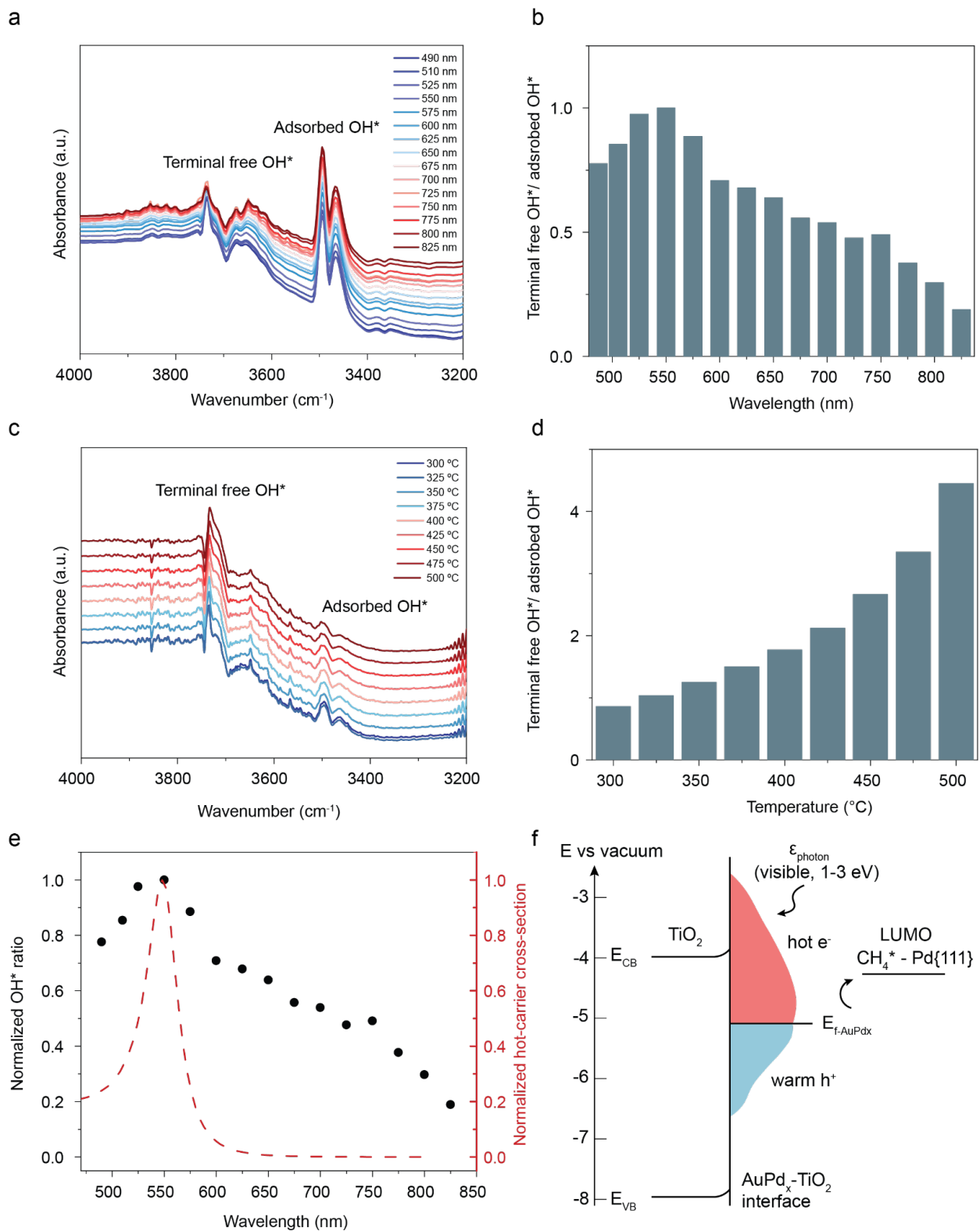

**Figure 4: Wavelength dependency and temperature dependency of the intermediates.** (a) *In-situ* DRIFTS under light illumination at different wavelengths and (b) the ratio between OH groups on the AuPd$_{0.05}$/TiO$_2$ catalyst. (c) *In-situ* DRIFTS under thermocatalytic conditions at different temperatures and (d) the ratio between OH groups on the AuPd$_{0.05}$/TiO$_2$ catalyst. (e) Normalized OH group ratio overlaid with the normalized hot-carrier cross section (f). Band diagram of the AuPd/TiO$_2$ catalyst.

**Quantum Mechanistic Calculations of the Methane Coupling**.

To further substantiate the mechanistic picture implied by the DRIFTS and hot-carrier analyses, we performed ground-state density functional theory (DFT) calculations and high-accuracy embedded correlated wavefunction (ECW)-based quantum mechanical calculations.[58] These calculations resolve the intrinsic energy landscape governing methane coupling on AuPd/TiO$_2$.(Fig. 5) The optimized intermediates and transition states along the minimum-energy pathway (MEP) reveal that C–C coupling formation is initiated by the coupling of two surface-bound methyl species ($CH_3^* + CH_3^* \rightarrow C_2H_6^*$). The MEP were optimized using plane-wave (PW) DFT with the climbing-image nudged elastic band (CI-NEB) method, employing the Perdew–Burke–Ernzerhof functional with Grimme's D3 dispersion correction and Becke–Johnson damping (i.e., PW-PBE-D3BJ) (see Methods for the details). The resulting activation barrier is 2.07 eV, making this step a rate-determining step (RDS) of the overall network (Fig. 5a,b, Supplementary Fig. 10). Subsequent C–C bond extension through the reaction of $C_2H_5^*$ with an additional $CH_3^*$ fragment proceeds with a comparable barrier (Fig. 5c, d). Hence, controlling the early-stage $CH_3^*$ recombination step is essential for directing selectivity toward multicarbon products.

Integrating these ground-state energetics with our experimental observations, we propose a unified surface mechanism in which hydrocarbon fragments generated by C–H activation coexist with molecular oxygen supplied through N$_2$O decomposition (Fig. 5e). The resulting competition between C–C coupling and deep oxidation is strongly modulated under plasmon excitation. Rather than simply injecting electrons into the oxide, hot-carrier excitation alters the effective surface oxygen coverage by transiently modifying O*/OH* binding and population. The reduced availability of reactive oxygen species suppresses overoxidation pathways, allowing hydrocarbon intermediates to persist long enough for C–C coupling to outcompete C=O formation.

To facilitate high-accuracy understanding of the $C_2H_6^*$ formation pathways on Pd (Fig. 5f), beyond-DFT investigations were carried out by calculating both the ground state (denoted as S0) and the localized surface plasmon resonance (LSPR)-accessible excited states (denoted as S1 to S9) along the reaction coordinate by the multi-state quantum embedded restricted active space second-order perturbation (emb-MS-RASPT2) theory (see Supplementary Fig. 11,12 and Methods for the details). The ground-state ECW energy refined the thermal barrier of C-C bond formation to 2.7 eV. This is significantly higher than the DFT-level PBE-D3BJ barrier due to the well-known electron delocalization error in the approximated exchange-correlation density functional, which over-stabilizes the transition state. Upon illumination, the electronically excited states become accessible, and an initial 1.4 eV S0-to-S1 excitation at the adsorbed local minimum was induced, which is then able to relax the structure to a new S1 local minimum at a reaction coordinate of 0.6 Å. This minimum can then be excited further by absorbing a photon energy of 1.5 eV, reaching the S9 state. The system can then evolve on the S9 state, and quickly hop to the nearby states non-adiabatically due to the strong couplings between a manifold of excited states, leaving only ~0.7 eV remaining to reach the accessible transition state on the S1 state. Therefore, our excited-state reactive trajectory has a reduced activation barrier of only ~0.7 eV, rendering the C–C coupling step kinetically accessible under experimental conditions.

The simplified schematic in Figure 5g highlights that, relative to the thermodynamic reference of methane combustion (red line), the barriers for $C_2$ and $C_3$ formation remain accessible where oxidative pathways are kinetically suppressed. (See Supplementary Fig. 13 for other possible pathways for $C_2H_4$ formation.) Taken together, these calculations provide a framework that complements our plasmon-mediated DRIFTS results, supporting a picture in which hot-carrier-driven redistribution of surface adsorbates shifts the reaction landscape away from overoxidation and toward selective C–C bond formation.

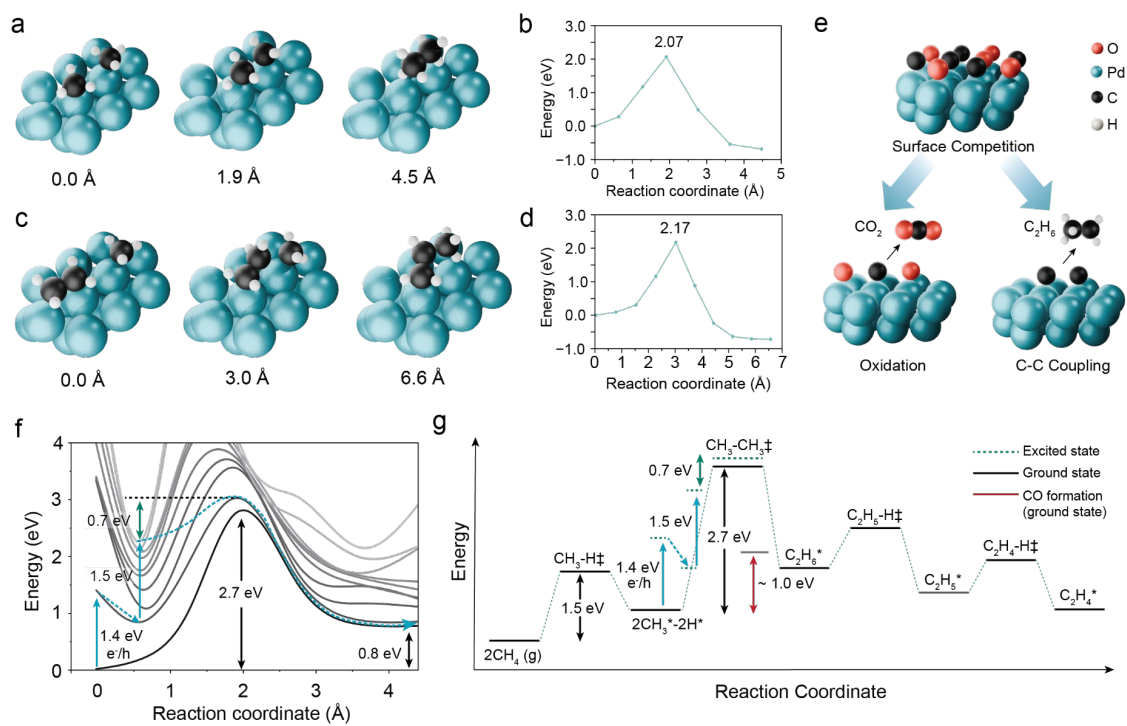

**Figure 5. Reaction pathway for methane coupling derived from quantum mechanical calculations.** (a,c) The isometric views of the optimized critical structures on the ground-state minimum-energy paths (MEPs) with the corresponding reaction coordinates (in Å) for $C_2H_6$*(a) and $C_3H_8$* (c) formation on a palladium (111) surface. (b,d) The PW-DFT relative energies along the MEPs for the surface $CH_3$* + $CH_3$* → $C_2H_6$* (b) and $C_2H_5$* + $CH_3$* → $C_3H_8$* (d) coupling reaction, where each of them serves as the rate-determining step in the reaction network. (e) Schematic illustration of our proposed surface mechanism for steering methane coupling pathways with a competition between C-C coupling and oxidation routes. (f) Reaction pathways on the ground-state and excited-state potential-energy surfaces, projected along the C-C bond formation coordinate, for $C_2H_6$* formation on the Pd surface. Calculations were performed using the emb-MS-RASPT2 method for all states. Refer to the ECW subsection in the Methods

section. The blue vertical arrows represent excitation processes, while the curved dashed arrow shows a semi-quantitative reactive trajectory under photocatalytic conditions. (g) Simplified schematic of the reaction pathways through methane coupling towards $C_2$ product formation. The red line shows the reference of overoxidation (methane combustion reaction) from previously reported literature.

**Conclusions**

This work demonstrates a plasmonic photocatalytic route for the simultaneous mitigation and upgrading of $CH_4$ and $N_2O$ under mild conditions, offering a low-carbon alternative to energy-intensive thermochemical processes. By tuning AuPd alloying on $TiO_2$, we identify a composition that couples efficient light absorption with C–C coupling. Photocatalytic measurements reveal that only the optimally alloyed $AuPd_{0.05}/TiO_2$ catalyst produces $C_2$ and $C_3$ hydrocarbons, while thermocatalytic conditions or nonoptimal alloy ratios yield mainly $CO_2$. *In-situ* DRIFTS and hot-carrier modeling indicate that plasmon-generated carriers at the metal–oxide interface modulate OH* coverage and shift the reaction pathway from overoxidation toward selective oxidative coupling. Ground state DFT calculations and excited-state ECW analysis further show that $CH_3$* recombination is one of the RDS that govern the $C_2$ and $C_3$ product distribution, providing a mechanistic basis for controlling selectivity.

Together, these findings outline a generalizable strategy for leveraging plasmon-mediated interfacial chemistry to redirect oxidative transformations of small molecules. To advance this approach toward practical deployment, further work is needed to improve overall photon-to-product efficiency, optimize light–matter coupling at the reactor and catalyst levels, and reduce the optical power requirements under realistic illumination conditions. Catalyst-level optimizations may include engineering the support to enhance charge extraction and thermal management, tuning particle morphology and loading to maximize plasmonic utilization, and replacing Au with earth-abundant plasmonic or alloyed alternatives while preserving hot-carrier selectivity. At the system level, integrating structured supports or photonic architectures could improve light delivery and scalability. This approach reveals design principles for catalytic interfaces that balance oxidation and coupling pathways, thereby positioning plasmon-driven photocatalysis as a promising platform for translating greenhouse-gas valorization from proof-of-concept studies toward energy-efficient, industrially relevant photochemical processes.

**Methods**
**Materials preparation**
$AuPd_x/TiO_2$ (x = 0.01, 0.05, 0.1, 0.2, 0.3) and $Pd/TiO_2$ catalysts were synthesized following a previously reported co-deposition–precipitation protocol, using urea as a reducing and precipitating agent. Commercial titania (brand) was used as the support, while $HAuCl_4 \cdot 3H_2O$ and $Pd(aca)_2$ served as the gold and palladium precursors to achieve a nominal 3 wt% metal loading. Prior to synthesis, the $TiO_2$ powder was dried in air at 100 °C for approximately 24 h to remove surface moisture and improve precursor–support interactions. Subsequently, 500 mg of the dried $TiO_2$ was dispersed into 37 mL of an aqueous solution containing pre-dissolved $HAuCl_4$, $Pd(aca)_2$, and urea (0.42 M). The suspension was heated to 85 °C using an oil bath and maintained under continuous magnetic stirring in a round-bottom flask equipped with a condenser. The reaction proceeded overnight (>16 h). After the reaction, the mixture was centrifuged, washed three times with deionized water, and the resulting solid was collected. The recovered material was then dried under vacuum at 115 °C for 2 h and stored in a vacuum desiccator at room temperature, protected from light. The as-prepared catalyst was subsequently reduced in a Harrick cell under a flowing $H_2$ atmosphere at 400 °C for 1 h, followed by annealing in Ar at the same temperature for an additional hour to eliminate residual hydrides and prevent palladium hydride formation prior to catalytic testing.

**Photocatalytic reaction**

The catalyst bed is generally pressed as a cylinder-shaped pellet in a Harrick cell with dimensions of ~ 6mm in diameter, >1 mm in height, for 15 mg. A custom-built gas-feeding system in Dionne Lab is employed [49] to introduce the mixture with the combination of $CH_4$, $N_2O$, and Ar into the Harrick cell for thermocatalytic and photocatalytic measurements. In a typical methane coupling to high-value hydrocarbon experiment, 20 sccm $CH_4$, 10 sccm Ar, and 2 sccm $N_2O$ were constantly flowed into a Harrick cell, and the mixture gas was finally flowed into a gas chromatograph (SRI instrument, 8610 GC) equipped with both thermal conductance detector(TCD) and flame ion detector(FID) using Ar as the carrier gas. Optical illumination was performed with an NKT superK laser equipped with a tunable Varia bandpass filter (bandwidth was set as 440 nm for white illumination and 50 nm for the wavelength-dependent measurements). Surface temperatures of the catalyst beds were monitored using an infrared thermography camera with a temporal resolution of 30 FPS and a spatial resolution of ~ 100 μm to monitor the max temperature on the catalyst bed.

**Materials Characterization**
**a) Diffuse Reflectance UV-Vis spectroscopy**

Diffuse reflectance spectra were collected using an Agilent Cary 5000 UV-Vis-NIR spectrophotometer equipped with a Praying Mantis™ diffuse reflectance accessory (Harrick Scientific, DRP-VA) to enable measurement in reflection mode. After the high-temperature reduction treatment, the sample was taken out and transferred to the Praying Mantis holder. Spectra were recorded over the 400–800 nm range at a scan rate of 600 nm/min. Magnesium oxide powder (MgO, Sigma-Aldrich, #342793) was used as the white reference for background correction.

**b) Monte-Carlo simulations**

Photon transport within colloidal systems containing AuPd alloy and $TiO_2$ nanoparticles dispersed in a gaseous environment was modeled using a custom Monte Carlo simulation implemented in Python. The simulation captures the nanoscale optical behavior of the catalyst bed by incorporating Mie theory-based scattering and absorption properties of spherical nanoparticles.

To model normal incident light illumination, photons were initialized uniformly on the surface of the catalyst bed, with velocity vectors normal to the surface. No reflection was assumed at the

gas–catalyst interface. The photon path length between consecutive scattering or absorption events was sampled from an exponential distribution with probability density function $\alpha \cdot e^{-\alpha x}$, where the extinction coefficient $\alpha = \alpha_s + \alpha_a$ is the sum of the scattering coefficient $\alpha_s$ and the absorption coefficient $\alpha_a$.

Both AuPd and TiO$_2$ nanoparticles were approximated as spheres. The total scattering coefficient α$_s$ was computed as the sum of the scattering contributions from AuPd and TiO$_2$ particles, each calculated as the product of their Mie-derived scattering cross sections and corresponding number densities. The absorption coefficient $\alpha_a$ was taken as the product of the Mie-derived absorption cross section and number density of AuPd nanoparticles only, as TiO$_2$ was assumed to be non-absorbing in the considered spectral range.

Upon collision with a nanoparticle, a fraction $\alpha_a/(\alpha_s + \alpha_a)$ of the photon's energy was absorbed, and the remaining portion was scattered into a new direction based on an unpolarized dipole far-field angular distribution. Photon tracking continued until the photon's energy fell below a defined threshold (assumed fully absorbed) or it exited the catalyst bed.

To account for particle size effects, AuPd nanoparticles were assigned radii sampled from a realistic size distribution, and Mie optical properties were computed accordingly. For TiO$_2$ particles, a size-averaged scattering coefficient was used by integrating over particle radii from 420 to 500 nm to smooth spectral features. The catalyst absorption spectrum was estimated at each wavelength by calculating the ratio of absorbed photons to a total of 100,000 incident photons.

### c) *In-situ* Diffuse Reflective Infrared Fourier-Transform Spectroscopy

*In-situ* diffuse reflectance infrared Fourier-transform spectroscopy (DRIFTS) measurements were performed under identical gas flow, heating, and light irradiation conditions as those used for the photocatalytic reactions. The experiments employed a Praying Mantis DRIFTS setup (Harrick) coupled to a Thermo Fisher Nicolet IS50 spectrometer. Prior to data acquisition, the optical path of the DRIFTS cell was purged overnight with clean, dry air. Infrared spectra were recorded using a liquid-nitrogen-cooled HgCdTe (MCT) detector. A reference spectrum was obtained by averaging 16 scans before initiating the *in-situ* reaction. All spectra are shown in absorbance form, calculated as log(I/I$_0$), where I$_0$ and I denote the background and reaction spectra, respectively.

### d) Hot-carrier calculations

Hot-carrier generation in AuPd nanoparticles was modeled using the semiclassical spherical formalism developed by the Govorov group, including both Landau damping and interband transitions. Because the AuPd nanoparticles in the catalyst were 3 to 6 nm in diameter, they were treated within the quasistatic limit and approximated as spheres. The high-energy hot-carrier generation rate arising from Landau damping was evaluated using the symmetry-reduced spherical expression: $Rate_{high\,energy,sphare} \approx \frac{2}{\pi^2} \cdot \frac{e^2 E_F^2}{\hbar \omega^3} \cdot \frac{4\pi R_0^3}{3} \cdot \frac{\varepsilon_0^2}{|\varepsilon_{metal}+2\varepsilon_0|^2} \cdot \frac{I_0}{c}$, where $e$ is the electron charge, $E_F$ is the Fermi energy of AuPd, $\omega$ is the photon angular frequency, $R_0$ is

the nanoparticle radius, $\varepsilon_0$ is the permittivity of the surrounding medium, $\varepsilon_{metal}$ is the complex dielectric function of AuPd, $I_0$ is the incident light intensity, and $c$ is the speed of light. The term $\frac{4\pi R_0^3}{3}$ accounts for the nanoparticle volume, and the factor $\frac{\varepsilon_0^2}{|\varepsilon_{metal}+2\varepsilon_0|^2}$ represents the quasistatic dipolar field enhancement at the particle surface that governs the strength of Landau damping.

Interband contributions were computed using the corresponding spherical quasistatic expression: $R_{inter}^{(sphere)}(\omega) \approx \frac{\varepsilon_{inter}^{''}(\omega)}{\hbar c} \cdot \left|\frac{3\varepsilon_{env}}{\varepsilon_m(\omega)+2\varepsilon_{env}}\right|^2 \cdot \frac{4\pi R^3}{3} I_0$, where $\varepsilon_{inter}^{''}(\omega)$ is the imaginary component of the interband dielectric response, $\varepsilon_{env}$ is the permittivity of the surrounding medium, and $\varepsilon_m(\omega)$ is the full metal dielectric function. In this expression, $\varepsilon_{inter}^{''}(\omega)$ captures direct d-to-sp transitions, while the factor $\left|\frac{3\varepsilon_{env}}{\varepsilon_m(\omega)+2\varepsilon_{env}}\right|^2$ represents the enhancement of the internal electric field that drives interband excitation.

The total hot-carrier generation rate for each particle radius was obtained by summing the Landau and interband terms. To account for polydispersity, the radius-dependent hot-carrier rates were evaluated over the experimentally measured AuPd size distribution and averaged to obtain the ensemble response. All calculations, including dielectric interpolation and wavelength-resolved rate evaluation, were performed using a custom Python code developed in-house.

### e) HAADF-STEM/EDS
TEM and STEM imaging were carried out at the Stanford Nano Shared Facilities (SNSF). TEM imaging was performed on an FEI Titan environmental TEM operated at 300 kV and equipped with a Gatan OneView camera. High-angle annular dark-field STEM (HAADF-STEM) imaging and STEM–EDX elemental mapping were conducted on a Thermo Fisher Spectra TEM operated at 300 kV.

### f) Tomography
The tomographic reconstructions in this work are performed using the open-source Python package quantEM [59]. Using implicit neural representations with a neural network as a model regularizer, we jointly optimize the volume and the relative orientations of each tilt image, allowing for the most accurate reconstruction possible. This framework also compensates for the missing wedge, minimizing streaking artifacts in the volume. We also perform background subtraction using Bernstein polynomial fitting and a coarse cross-correlation alignment prior to reconstruction. The reconstruction ran for 150 iterations, where the volume and alignment parameters were fully converged.

### g) XAS
*Ex-situ* XAS spectra at the Pd K-edge (24,350 eV) were collected at beamline 7-BM (QAS) of NSLS-II under ambient conditions. For each measurement, the catalyst powder was pressed into a uniform pellet and sealed with Kapton tape. The XAS spectra were acquired in both transmission and fluorescence modes simultaneously. A Pd metal foil was measured alongside the samples and used as the energy reference for calibrating any energy shifts. All XAS data processing, including energy calibration, background subtraction, and normalization, was performed using the Demeter software package.[60]

**Quantum mechanical calculations**

**a) Periodic Plane-wave Density-Functional Theory (PW-DFT).**

Periodic PW-DFT calculations were performed using the Vienna Ab Initio Simulation Package (VASP) version 5.4[61] with the Perdew-Burke-Ernzerhof (PBE)[62] exchange-correlation (XC) functional. Grimme's D3 dispersion corrections[63] were added with the Becke-Johnson (BJ) damping function to correct for the deficiencies in non-covalent interactions in PBE. The projector augmented-wave (PAW) method was used, in which we explicitly treat the 1s electron for H, the 2s and 2p electrons for C and O, and the 4d and 5s electrons for Pd. A plane-wave (PW) basis kinetic energy cutoff of 500 eV with a Methfessel-Paxton smearing width of 0.2 eV was used during the structure optimizations. A $1 \times 1 \times 1$ k-point mesh was used to perform the electronic integrations during the geometry optimizations, and a refined $5 \times 5 \times 1$ k-point mesh was applied for the final DFT-level single-point energies. An energy convergence threshold of $10^{-6}$ eV was used for electronic self-consistent field (SCF) convergence. A four-layer 100-atom Pd(111) slab was optimized to model the surface of the nanoparticle. Between the images in the direction that is normal to the surface, a 19-Å-thick vacuum layer was added. A dipole correction along this direction was applied to eliminate unphysical interactions between periodic images. The ground-state minimum-energy paths (MEPs) were optimized by using the climbing image nudged elastic band (CI-NEB) method. The convergence threshold for the maximum of the atomic forces on the images of the NEB was set to be 0.03 eV/Å. The reaction coordinates ($s$ in Å) used for representing the converged CI-NEB paths is defined as $s$: $s(n) = \sum_{j=1}^{n} |r_{j+1} - r_j|$, where $n$ is the image index, and $r_j$ is the atomic cartesian coordinate vector of the *j*-th image.

**b) Embedded Correlated Wavefunction (ECW) Theory.**

In the ECW theory,[64,65] the whole periodic system is partitioned into a cluster and its environment. The interaction between the cluster and environment is approximated by a quantum embedding potential $V_{emb}(r)$, which is optimized using periodic PW-DFT. A 16-atom cluster was carved out from the periodic Pd slab (Supplementary Fig. 11). The PBE XC functional and the PW basis were used to calculate the embedding potential via the optimized effective potential method.

The final ECW energy for every image, $E^{ECW}$, is calculated as $E^{ECW} = E_{tot}^{PW-DFT} + (E_{clst}^{CW} - E_{clst}^{DFT})$, where $E_{tot}^{PW-DFT}$ is the total PW-PBE energy (without the D3-BJ correction) for the periodic system. The energy correction term, i.e., $E_{clst}^{CW} - E_{clst}^{DFT}$, is computed using the 16-atom carved cluster with atom-centered Gaussian-type basis in the presence of the embedding potential. The adsorbate was added onto the cluster with its geometry (internal coordinates) fixed at the corresponding PW-PBE-D3BJ optimized structures along the MEPs. $E_{clst}^{DFT}$ was computed with the PBE XC functional (without the D3-BJ correction) in the presence of the embedding potential. The basis sets used for the cluster calculations were the atomic natural orbital-relativistic correlation consistent valence triple zeta polarization (ANO-RCC-VTZP) basis for adsorbates, and the 2nd-generation Karlsruhe polarization consistent triple-zeta valence (def2-TZVP) basis for Pd.

For $E_{clst}^{CW}$, we performed embedded restricted active space second-order perturbation theory (emb-RASPT2). The ground-state wavefunctions were optimized using the single-state embedded restricted active space self-consistent field (emb-RASSCF) method. For surface-bound $CH_3$ recombination, we partitioned a large active space of (16e, 16o) using the four-particle four-hole embedded restricted active space self-consistent field (emb-RASSCF) method: (14e,7o) in RAS1, including six $CH_3$ $\sigma$ orbitals (three per $CH_3$ group, derived from C $2s$ and $2p$ atomic orbitals) and one Pd $4d_{z^2}$ – C $2p_z$ $\sigma$ orbital; (2e, 2o) in RAS2, which includes one Pd $4d_{z^2}$ – C $2p_z$ $\sigma$ orbital and the corresponding $\sigma^*$ orbital; and (0e,7o) in RAS3, including six CH $\sigma^*$ orbitals and one Pd $4d_{z^2}$ – C $2p_z$ $\sigma$ orbital. (See Supplementary Fig. 12 for images of the optimized active space.)

An ionization-potential electron-affinity (IPEA) shift of 0.25 hartree was used in the emb-RASPT2 calculations, recovering dynamical correlations, which were performed with the single-state emb-RASSCF wavefunctions. The embedded state-averaged restricted active space self-consistent field (emb-SA-RASSCF) method was used to optimize the excited-state wavefunctions while always including the ground-state wavefunction in the average and an equal weight for each root in the state optimization. The excitation energies were computed by emb-multistate-RASPT2 with the wavefunctions optimized by emb-SA-RASSCF. An imaginary shift of 0.2 hartree was applied to avoid intruder states. The excitation energies were added onto the ground-state single-state emb-RASPT2 energy to obtain the excited-state PESs. An in-house-modified Molcas program[66] was used to perform all the ECW calculations on the cluster.

**Acknowledgement**
All authors at Stanford acknowledge the CryoEM support from the Office of Basic Energy Sciences, U.S. Department of Energy, Division of Materials Science and Engineering (DE-AC02-76SF00515). S.L. and J.A.D. acknowledge the support from the U.S. Department of Energy Office of Science National Quantum Information Science Research Centers as part of the Q-NEXT center. S.L., L.Y., and J.A.D. acknowledge financial support from the National Research Foundation of Korea (NRF) grant funded by the Korean Government (Ministry of Science and ICT) (No. RS-2024-00421181). S.L., L.Y., and J.A.D. acknowledge the use and support of the Stanford Nano Shared Facilities (SNSF), supported by the National Science Foundation under award ECCS-2026822. This research used resources of the National Energy Research Scientific Computing Center, a DOE Office of 360 Science User Facility supported by the Office of Science of the U.S. Department of Energy under Contract No.361 DE-AC02-05CH11231, using AI4Sci@NERSC NERSC award NERSC DDR-ERCAP0038157. This work utilized beamline 7-BM (QAS) at the National Synchrotron Light Source II (NSLS-II), which is a US Department of Energy Office of Science User Facilities. J.L.B. acknowledges the supercomputing cluster provided by Boston College, as well as financial support from the American Chemical Society Petroleum Research Fund (PRF number 65744-DNI6) and, in part,

**Reference**

1. Arias, P. *et al.* Climate Change 2021: the physical science basis. Contribution of Working Group I to the Sixth Assessment Report of the Intergovernmental Panel on Climate Change; technical summary. (2021).

2. Etminan, M., Myhre, G., Highwood, E. J. & Shine, K. P. Radiative forcing of carbon dioxide, methane, and nitrous oxide: A significant revision of the methane radiative forcing. *Geophys. Res. Lett.* **43**, (2016).

3. Li, Q., Ouyang, Y., Li, H., Wang, L. & Zeng, J. Photocatalytic conversion of methane: Recent advancements and prospects. *Angew. Chem. Weinheim Bergstr. Ger.* **134**, (2022).

4. Song, H., Meng, X., Wang, Z.-J., Liu, H. & Ye, J. Solar-energy-mediated methane conversion. *Joule* **3**, 1606–1636 (2019).

5. Van Der Laan, G. P. & Beenackers, A. A. C. M. Kinetics and selectivity of the Fischer–tropsch synthesis: A literature review. *Catal. Rev. Sci. Eng.* **41**, 255–318 (1999).

6. Barelli, L., Bidini, G., Gallorini, F. & Servili, S. Hydrogen production through sorption-enhanced steam methane reforming and membrane technology: A review. *Energy (Oxf.)* **33**, 554–570 (2008).

7. Shukla, P. R. *et al.* Climate change 2022: Mitigation of climate change. *Contribution of working group III to the sixth assessment report of the Intergovernmental Panel on Climate Change* **10**, 9781009157926 (2022).

8. Farrell, B. L., Igenegbai, V. O. & Linic, S. A viewpoint on direct methane conversion to ethane and ethylene using oxidative coupling on solid catalysts. *ACS Catal.* **6**, 4340–4346 (2016).

9. Galadima, A. & Muraza, O. Revisiting the oxidative coupling of methane to ethylene in the golden period of shale gas: A review. *J. Ind. Eng. Chem.* **37**, 1–13 (2016).

10. Zhang, W. *et al.* High-performance photocatalytic nonoxidative conversion of methane to ethane and hydrogen by heteroatoms-engineered TiO2. *Nat. Commun.* **13**, 2806 (2022).

11. Parmon, V. N., Panov, G. I., Uriarte, A. & Noskov, A. S. Nitrous oxide in oxidation chemistry and catalysis: application and production. *Catal. Today* **100**, 115–131 (2005).

12. Yu, S. & Jain, P. Plasmonic photosynthesis of C1–C3 hydrocarbons from carbon dioxide assisted by an ionic liquid. *Nat. Commun.* **10**, (2019).

13. Yu, S., Wilson, A. J., Heo, J. & Jain, P. K. Plasmonic control of multi-electron transfer and C-C coupling in visible-light-driven CO2 reduction on Au nanoparticles. *Nano Lett.* **18**, 2189–2194 (2018).

14. Kumari, G., Zhang, X., Devasia, D., Heo, J. & Jain, P. K. Watching visible light-driven CO2 reduction on a plasmonic nanoparticle catalyst. *ACS Nano* **12**, 8330–8340 (2018).

15. Robatjazi, H. *et al.* Plasmon-driven carbon–fluorine (C(sp3)–F) bond activation with mechanistic insights into hot-carrier-mediated pathways. *Nat. Catal.* **3**, 564–573 (2020).

16. Swearer, D. F. *et al.* Heterometallic antenna−reactor complexes for photocatalysis. *Proc. Natl. Acad. Sci. U. S. A.* **113**, 8916–8920 (2016).

17. Yuan, L. *et al.* Plasmonic photocatalysis with chemically and spatially specific antenna-dual reactor complexes. *ACS Nano* **16**, 17365–17375 (2022).

18. Yuan, Y. *et al.* Steam methane reforming using a regenerable antenna–reactor plasmonic photocatalyst. *Nat. Catal.* **7**, 1339–1349 (2024).

19. Yuan, L. *et al.* Atmospheric-pressure ammonia synthesis on AuRu catalysts enabled by plasmon-controlled hydrogenation and nitrogen-species desorption. *Nat. Energy* **11**, 98–108 (2025).

20. Zhou, L. *et al.* Light-driven methane dry reforming with single atomic site antenna-reactor plasmonic photocatalysts. *Nat. Energy* **5**, 61–70 (2020).

21. Bayles, A. *et al.* Tailoring the aluminum nanocrystal surface oxide for all-aluminum-based antenna-reactor plasmonic photocatalysts. *Proc. Natl. Acad. Sci. U. S. A.* **121**,

e2321852121 (2024).

22. Lee, W. H. *et al.* Polymeric stabilization at the gas-liquid interface for durable solar hydrogen production from plastic waste. *Nat. Nanotechnol.* **20**, 1237–1246 (2025).

23. Yang, H. *et al.* Packing-induced selectivity switching in molecular nanoparticle photocatalysts for hydrogen and hydrogen peroxide production. *Nat. Nanotechnol.* **18**, 307–315 (2023).

24. Brongersma, M. L., Halas, N. J. & Nordlander, P. Plasmon-induced hot carrier science and technology. *Nat. Nanotechnol.* **10**, 25–34 (2015).

25. Zhu, M. & Li, Z. Floating solar hydrogen production. *Nat. Nanotechnol.* **18**, 698–699 (2023).

26. Teng, Z. *et al.* Asymmetric photooxidation of glycerol to hydroxypyruvic acid over Rb-Ir catalytic pairs on poly(heptazine imides). *Nat. Nanotechnol.* **20**, 815–824 (2025).

27. Zhang, J. *et al.* Gold-modified nanoporous silicon for photoelectrochemical regulation of intracellular condensates. *Nat. Nanotechnol.* **20**, 835–844 (2025).

28. Cheruvathoor Poulose, A. *et al.* Fast and selective reduction of nitroarenes under visible light with an earth-abundant plasmonic photocatalyst. *Nat. Nanotechnol.* **17**, 485–492 (2022).

29. Song, H. & Ye, J. Direct photocatalytic conversion of methane to value-added chemicals. *Trends Chem.* **4**, 1094–1105 (2022).

30. Wu, S. *et al.* Ga-doped and Pt-loaded porous TiO2-SiO2 for photocatalytic nonoxidative coupling of methane. *J. Am. Chem. Soc.* **141**, 6592–6600 (2019).

31. Li, L. *et al.* Synergistic effect on the photoactivation of the methane C-H bond over Ga(3+)-modified ETS-10. *Angew. Chem. Int. Ed Engl.* **51**, 4702–4706 (2012).

32. Yuliati, L., Hamajima, T., Hattori, T. & Yoshida, H. Highly dispersed Ce(III) species on silica and alumina as new photocatalysts for non-oxidative direct methane coupling. *Chem. Commun. (Camb.)* **0**, 4824–4826 (2005).

33. Meng, L. *et al.* Gold plasmon-induced photocatalytic dehydrogenative coupling of methane

to ethane on polar oxide surfaces. *Energy Environ. Sci.* **11**, 294–298 (2018).

34. Li, L. *et al.* Efficient sunlight-driven dehydrogenative coupling of methane to ethane over a Zn+-modified zeolite. *Angewandte Chemie International Edition* **50**, 8299–8303 (2011).

35. Jiang, W. *et al.* Pd-modified ZnO-Au enabling alkoxy intermediates formation and dehydrogenation for photocatalytic conversion of methane to ethylene. *J. Am. Chem. Soc.* **143**, 269–278 (2021).

36. Wang, C. *et al.* Synergy of Ag and AgBr in a pressurized flow reactor for selective photocatalytic oxidative coupling of methane. *ACS Catal.* **13**, 3768–3774 (2023).

37. Li, X. *et al.* PdCu nanoalloy decorated photocatalysts for efficient and selective oxidative coupling of methane in flow reactors. *Nat. Commun.* **14**, 6343 (2023).

38. Wu, M. *et al.* Photocatalytic oxidative coupling of methane to ethane using $CO_2$ as a soft oxidant over the Au/$TiO_2$-Vo nanosheets. *Angew. Chem. Int. Ed Engl.* **64**, e202414814 (2025).

39. Song, S. *et al.* A selective Au-ZnO/$TiO_2$ hybrid photocatalyst for oxidative coupling of methane to ethane with dioxygen. *Nat. Catal.* **4**, 1032–1042 (2021).

40. Song, H. *et al.* Integrating photochemical and photothermal effects for selective oxidative coupling of methane into C2+ hydrocarbons with multiple active sites. *Nat. Commun.* **16**, 2831 (2025).

41. Nie, W. *et al.* Photocatalytic oxidative coupling of methane to C3+ hydrocarbons via nanopore-confined microenvironments. *Nat. Energy* (2025) doi:10.1038/s41560-025-01834-5.

42. Peng, H. *et al.* Continuous flow photosynthesis of methanol from methane by plasmonic charge accumulation. *Nat. Commun.* **17**, 344 (2025).

43. Chen, Y. *et al.* Continuous flow system for highly efficient and durable photocatalytic oxidative coupling of methane. *J. Am. Chem. Soc.* **146**, 2465–2473 (2024).

44. Zhai, G. *et al.* Highly efficient, selective, and stable photocatalytic methane coupling to

ethane enabled by lattice oxygen looping. *Sci. Adv.* **10**, eado4390 (2024).

45. Li, X. *et al.* Efficient hole abstraction for highly selective oxidative coupling of methane by Au-sputtered TiO2 photocatalysts. *Nat. Energy* **8**, 1013–1022 (2023).

46. Kong, Y., Yang, C., Cai, Y., Mu, X. & Li, L. Photocatalytic aerobic conversion of methane. *Photocatalysis: Research and Potential* **1**, 10005–10005 (2024).

47. Arinaga, A. M., Ziegelski, M. C. & Marks, T. J. Alternative oxidants for the catalytic oxidative coupling of methane. *Angew. Chem. Int. Ed Engl.* **60**, 10502–10515 (2021).

48. Liu, H., Wei, Y., Caro, J. & Wang, H. Oxidative coupling of methane with high $C_2$ yield by using chlorinated perovskite $Ba_{0.5}Sr_{0.5}Fe_{0.2}co_{0.8}O_{3-\delta}$ as catalyst and $N_2O$ as oxidant. *ChemCatChem* **2**, 1539–1542 (2010).

49. Bourgeois, B. B. *et al.* Catalight -- an open source automated photocatalytic reactor package illustrated through plasmonic acetylene hydrogenation. *arXiv [physics.chem-ph]* (2025) doi:10.1021/acs.jpca.5c02883.

50. *IR Spectroscopy of Surface Water and Hydroxyl Species on Nanocrystalline TiO2 Films*.

51. Gholami, R., Alyani, M. & Smith, K. Deactivation of Pd catalysts by water during low temperature methane oxidation relevant to natural gas vehicle converters. *Catalysts* **5**, 561–594 (2015).

52. Baxter, R. J. & Hu, P. Insight into why the Langmuir–Hinshelwood mechanism is generally preferred. *J. Chem. Phys.* **116**, 4379–4381 (2002).

53. Doornkamp, C. & Ponec, V. The universal character of the Mars and Van Krevelen mechanism. *J. Mol. Catal. A Chem.* **162**, 19–32 (2000).

54. Yu, Y., Lundin, S.-T. B., Obata, K., Sarathy, S. M. & Takanabe, K. Improved homogeneous–heterogeneous kinetic mechanism using a Langmuir–Hinshelwood-based microkinetic model for high-pressure oxidative coupling of methane. *Ind. Eng. Chem. Res.* **62**, 5826–5838 (2023).

55. Zhang, H. *et al.* Unusual facet and co-catalyst effects in TiO2-based photocatalytic coupling

of methane. *Nat. Commun.* **15**, 4453 (2024).

56. Ageev, V. N. Desorption induced by electronic transitions. *Prog. Surf. Sci.* **47**, 55–203 (1994).

57. Smyczek, J. *et al.* Promoting role of isolated surface hydroxyls on selective dehydrogenation of 2-propanol to acetone over $Co_3O_4$ catalyst: A mechanistic study. *ACS Catal.* **15**, 19268–19280 (2025).

58. Libisch, F., Huang, C. & Carter, E. A. Embedded correlated wavefunction schemes: theory and applications. *Acc. Chem. Res.* **47**, 2768–2775 (2014).

59. Lim, C. *et al.* Missing wedge inpainting and joint alignment in electron tomography through implicit neural representations. *arXiv [eess.IV]* (2025) doi:10.48550/arXiv.2512.08113.

60. Ravel, B. & Newville, M. ATHENA, ARTEMIS, HEPHAESTUS: data analysis for X-ray absorption spectroscopy using IFEFFIT. *J. Synchrotron Radiat.* **12**, 537–541 (2005).

61. Kresse, G. & Furthmüller, J. Efficient iterative schemes for ab initio total-energy calculations using a plane-wave basis set. *Phys. Rev. B Condens. Matter* **54**, 11169–11186 (1996).

62. Perdew, J. P., Burke, K. & Ernzerhof, M. Generalized gradient approximation made simple. *Phys. Rev. Lett.* **77**, 3865–3868 (1996).

63. Grimme, S., Ehrlich, S. & Goerigk, L. Effect of the damping function in dispersion corrected density functional theory. *J. Comput. Chem.* **32**, 1456–1465 (2011).

64. Huang, C., Pavone, M. & Carter, E. A. Quantum mechanical embedding theory based on a unique embedding potential. *J. Chem. Phys.* **134**, 154110 (2011).

65. Martirez, J. M. P., Bao, J. L. & Carter, E. A. First-principles insights into plasmon-induced catalysis. *Annu. Rev. Phys. Chem.* **72**, 99–119 (2021).

66. Aquilante, F. *et al.* Molcas 8: New capabilities for multiconfigurational quantum chemical calculations across the periodic table: Molcas 8. *J. Comput. Chem.* **37**, 506–541 (2016).

**Supplementary Information**

**Plasmonic Photocatalysis Enables Selective Oxidative Coupling of Methane with Nitrous Oxide under Ambient Conditions**

Serin Lee†[1], Lin Yuan†[1], Elijah Begin[2], Dali Yang[3], Cedric Lim[1], Yirui Arlene Zhang[1], Lu Ma[3], Colin Ophus[1], Yi Cui[1], Junwei Lucas Bao[2] *, Jennifer A. Dionne[1,4] *

1. Department of Materials Science and Engineering, Stanford University School of Engineering, Stanford, CA, USA.
2. Department of Chemistry, Boston College, Chestnut Hill, MA, USA.
3. Brookhaven National Laboratory, Upton, NY, USA.
4. Department of Radiology, Stanford University School of Medicine, Stanford, CA, USA.

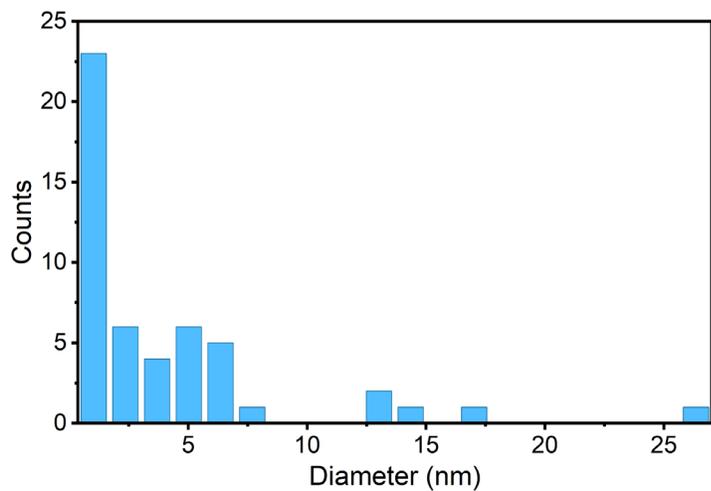

Supplementary Figure 1. Size distribution of AuPd nanoparticles on the support extracted from tomography reconstruction. The diameters were computed from the detected 50 nanoparticles in the reconstruction.

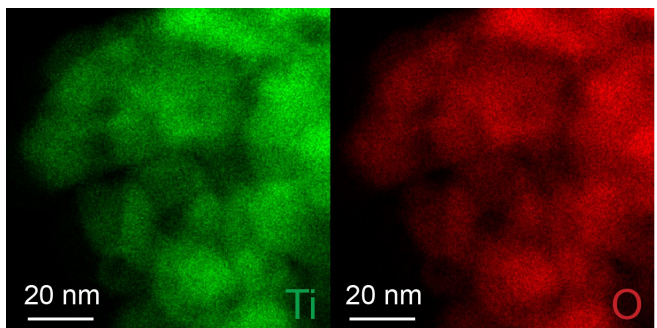

Supplementary Figure 2. EDS maps of Ti and O corresponding to the dashed region in Fig. 1d of the main text.

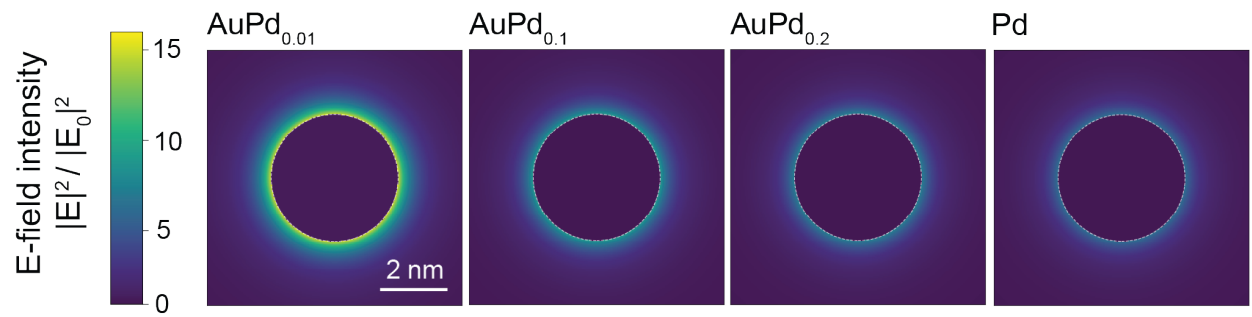

Supplementary Figure 3. Near-field simulation of the electric field at wavelength of 550 nm for $AuPd_{0.01}$, $AuPd_{0.1}$, $AuPd_{0.2}$, and Pd.

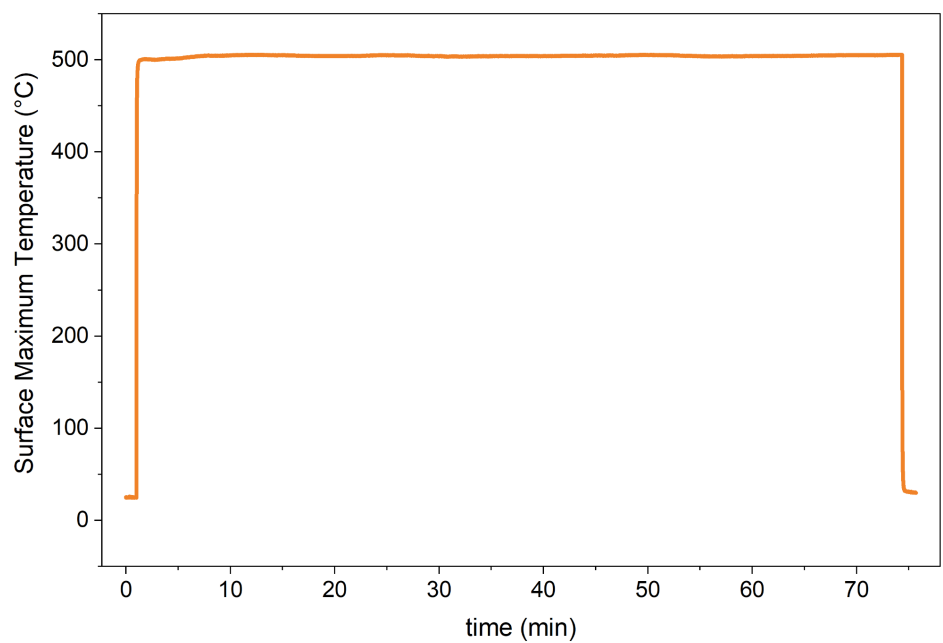

Supplementary Figure 4. Surface temperature measured using the infrared camera during the photocatalytic reaction.

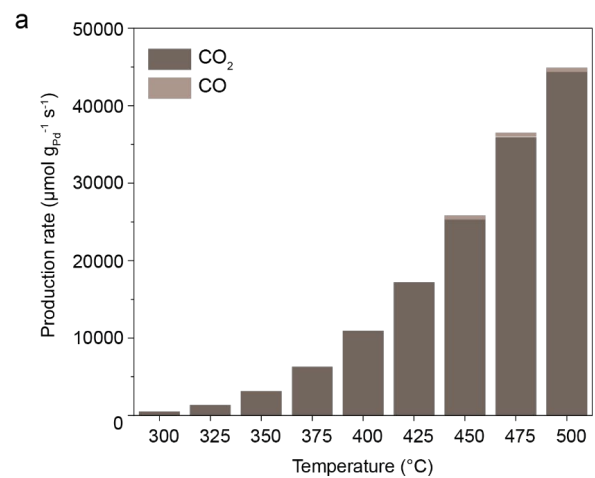
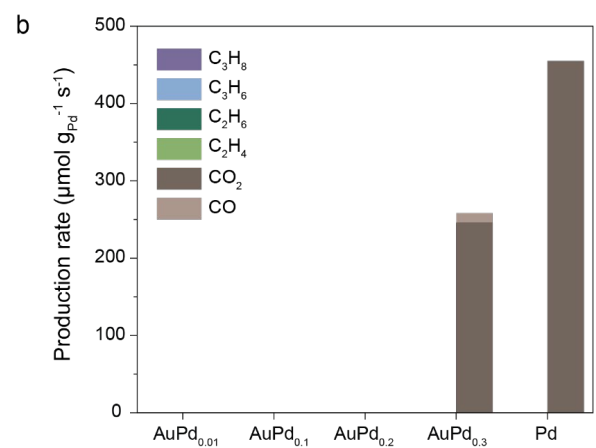
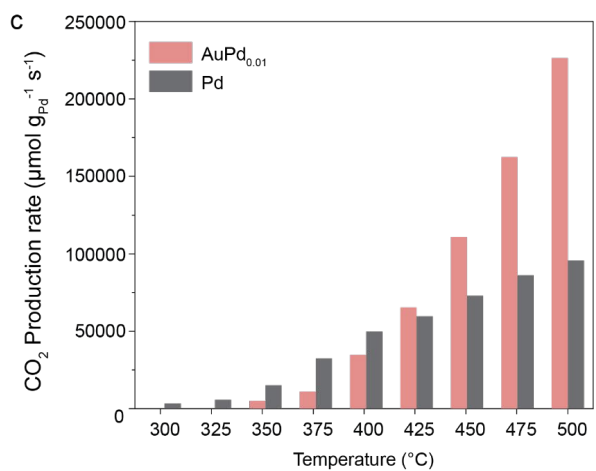

Supplementary Figure 5. (a) Heated photocatalytic methane conversion rates at different temperatures using 3 wt% AuPd$_{0.05}$/TiO$_2$ by illuminating 400-800 nm white light at 500 mW · cm$^{-2}$ under reactant gas concentrations of CH$_4$:N$_2$O:Ar =20:2:10. (b) Photocatalytic methane conversion rates using different ratios of Au and Pd by illuminating 400-800 nm white light at 500 mW · cm$^{-2}$ under reactant gas concentrations of CH$_4$:N$_2$O:Ar =20:2:10. (c) Thermocatalytic methane conversion rate using 3 wt% AuPd$_{0.01}$ /TiO$_2$ and Pd/TiO$_2$ catalysts under reactant gas concentration of CH$_4$:N$_2$O:Ar=20:2:10.

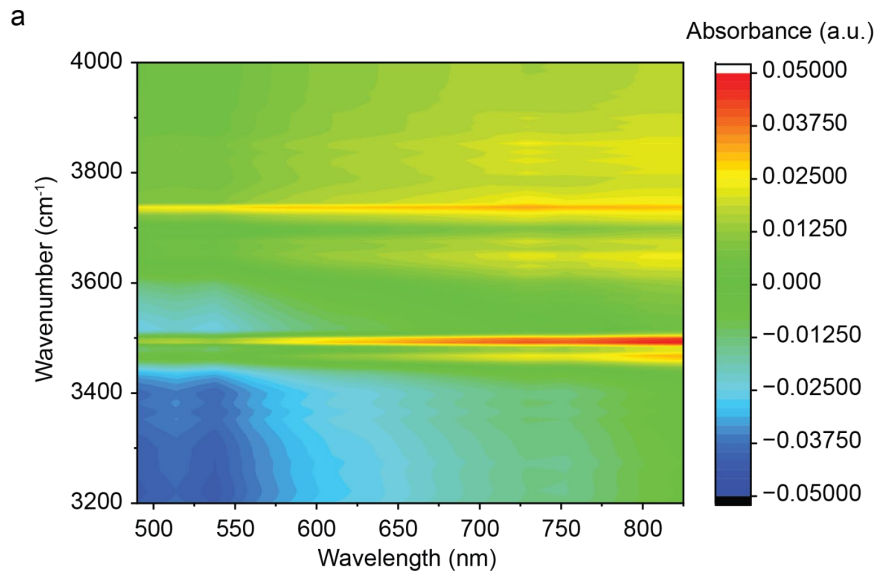

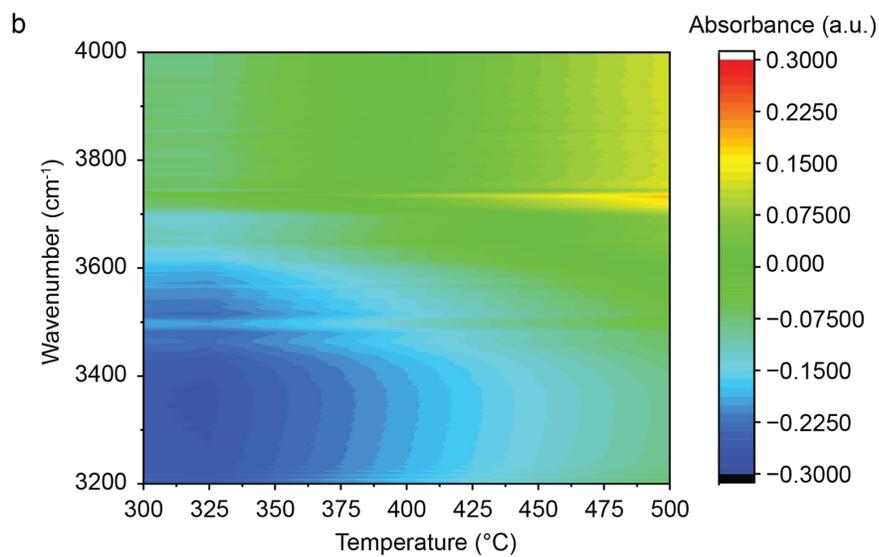

Supplementary Figure 6. Color-heatmap of *in-situ* DRIFTS under light illumination at different wavelengths (a) and under thermocatalytic conditions at different temperatures. (b)

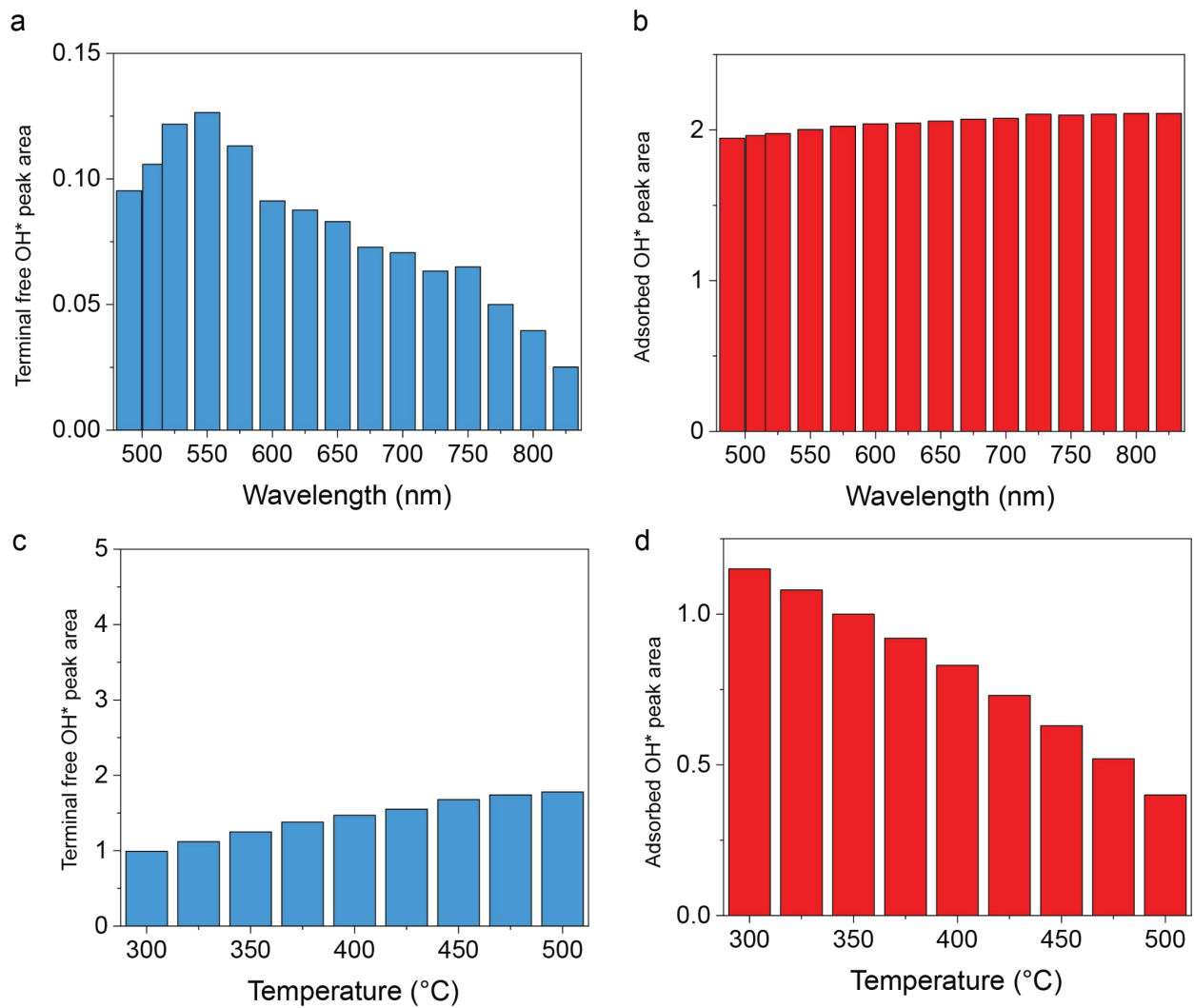

Supplementary Figure 7. Peak area corresponding to terminal free OH* and to adsorbed OH* from the main Figure 4 under photocatalysis condition at 100 mW by varying wavelength (a,b) and under thermocatalysis by varying temperature (c,d)

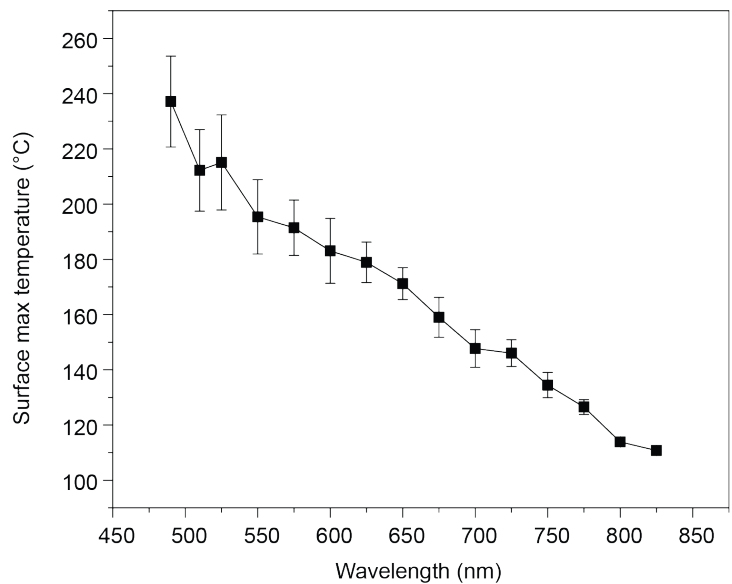

Supplementary Figure 8. Wavelength-dependent surface maximum temperature under 100 mW·cm$^{-1}$ during in-situ DRIFTS measurement.

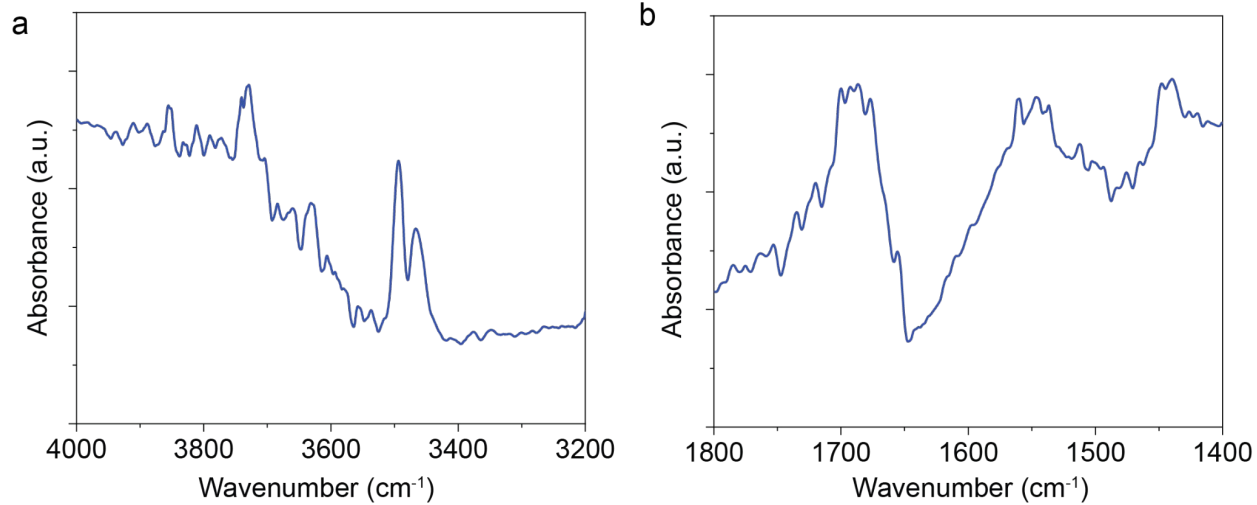

Supplementary Figure 9. *In-situ* DRIFTS on AuPd$_{0.05}$/TiO$_2$ under reactant gas concentration of CH$_4$:N$_2$O:Ar=20:2:10 by using 450 nm LED diode (600 mW · cm$^{-2}$) as a sole light source.

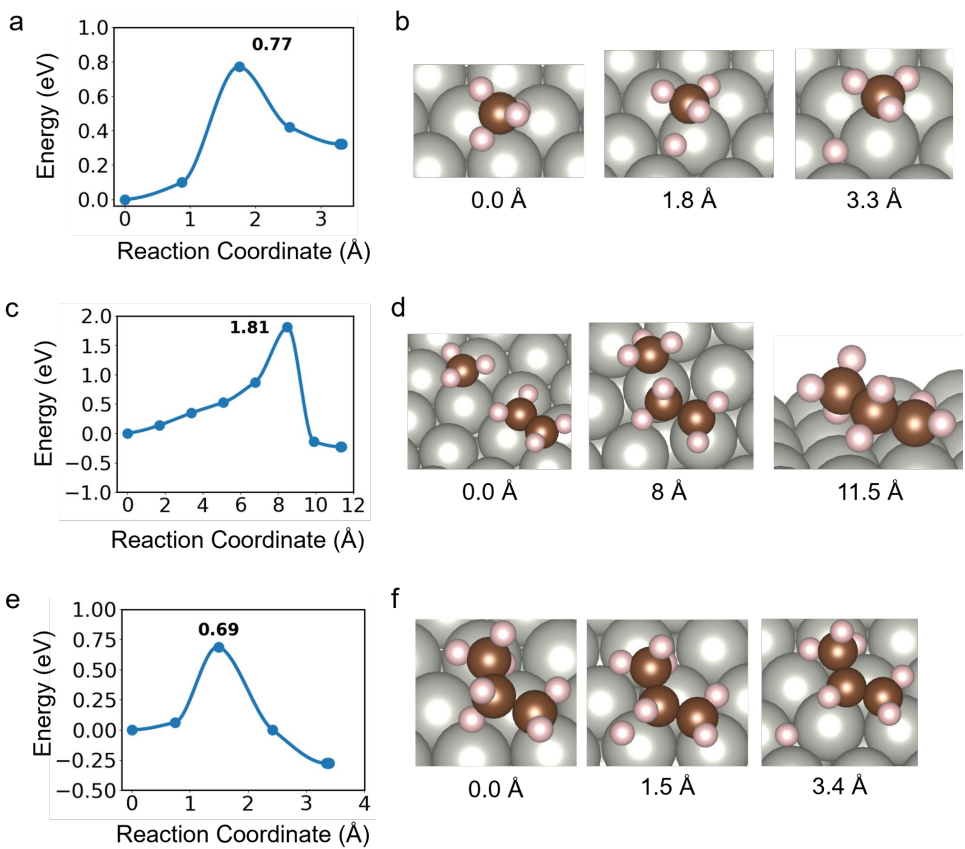

Supplementary Figure 10. Reaction pathway and visualization of surface-bound formation of $CH_4$(a,b), $C_3H_7$(c,d), and $C_3H_6$(e,f). Panels a, c, and e (left column) show the plane-wave DFT (PW-PBE-D3BJ) energies along the climbing image nudged elastic band (CI-NEB) minimum-energy paths (MEPs) as a function of the reaction coordinate (in Å). The corresponding panels b,d, and f (right column) display the structural models of the initial, transition, and final states along the MEP with the corresponding reaction coordinate. Gray, brown, and tan spheres represent Pd, C, and H atoms, respectively.

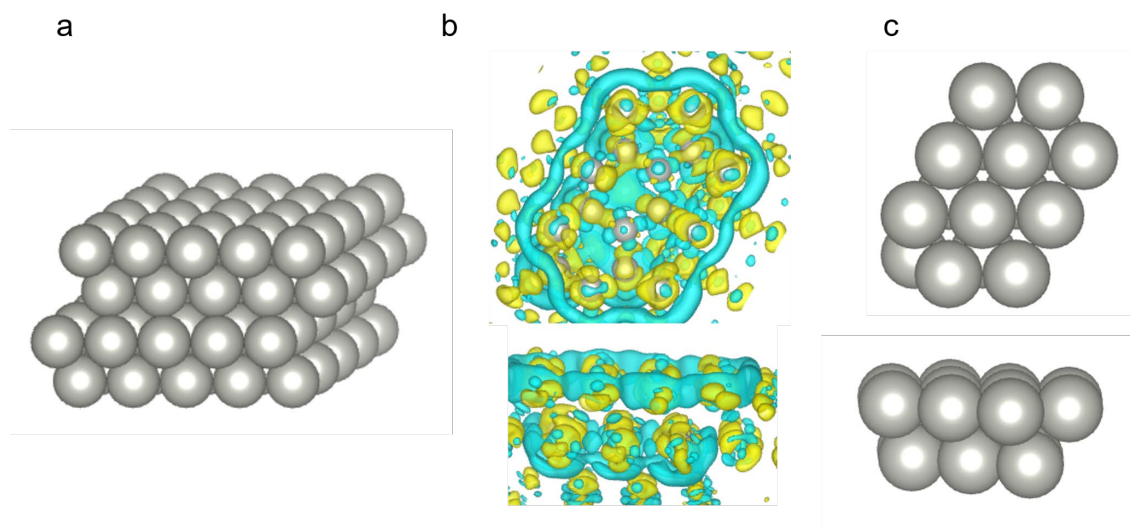

Supplementary Figure 11. The optimized quantum embedding potential used in the embedded correlated wavefunction (ECW) calculations. (a) Plane-wave DFT optimized Pd(111) slab geometry (which contains 100 Pd atoms per computational unit cell) that was used for optimizing the quantum embedding potential ($V_{emb}(r)$). (b) The top and side views of the optimized embedding potential isosurfaces, and (c) the carved $Pd_{16}$ cluster.

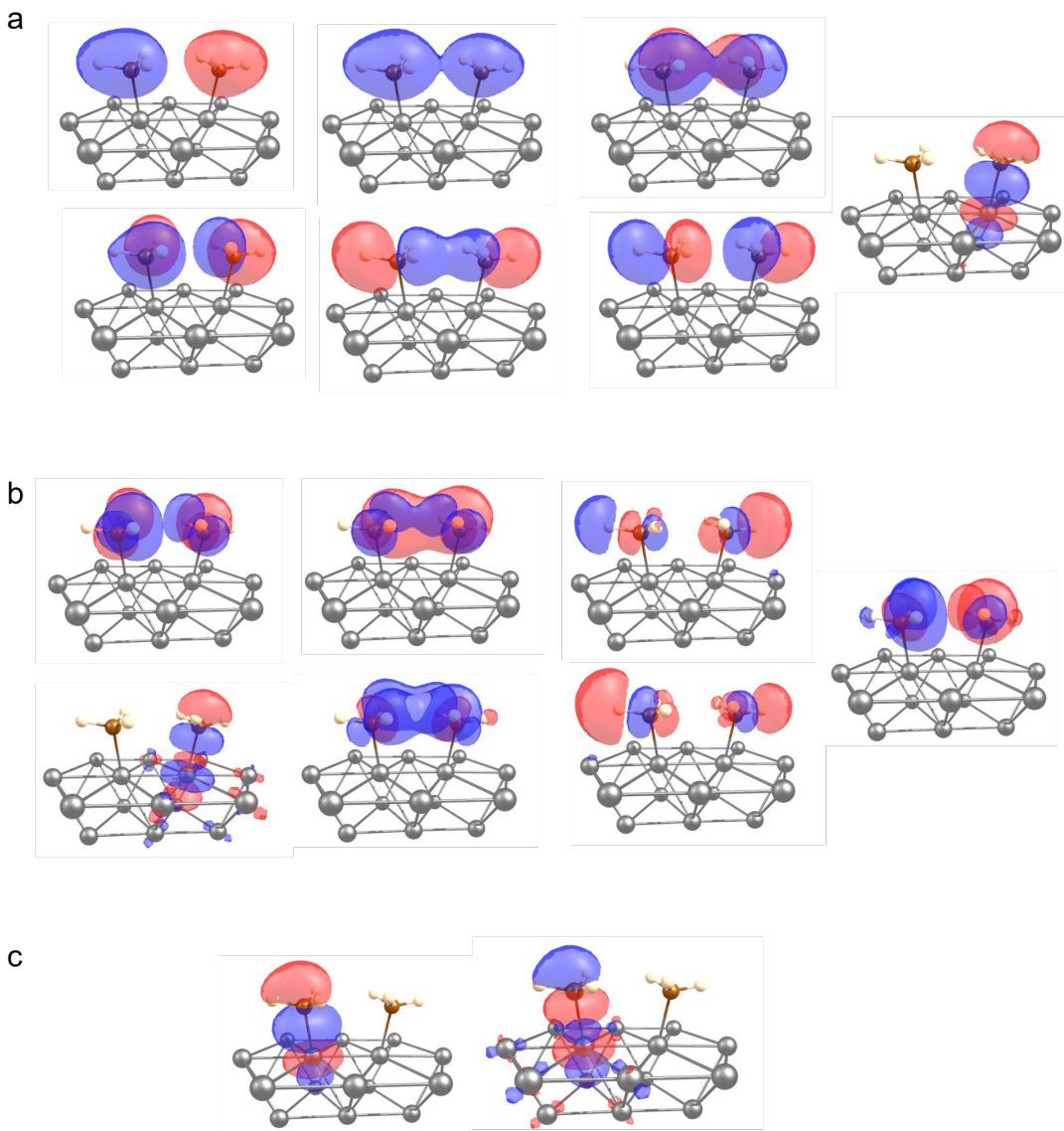

Supporting Figure 12. Optimized active space used for the ECW emb-RASPT2 calculations for surface-bound $C_2H_6$ formation visualized at the 0.0 Å reaction coordinate (the starting local-minimum structure). (a) The (14e,7o) RAS-1 space. (b) The (0e,7o) RAS-3 space. (c) The (2e,2o) RAS2 space.

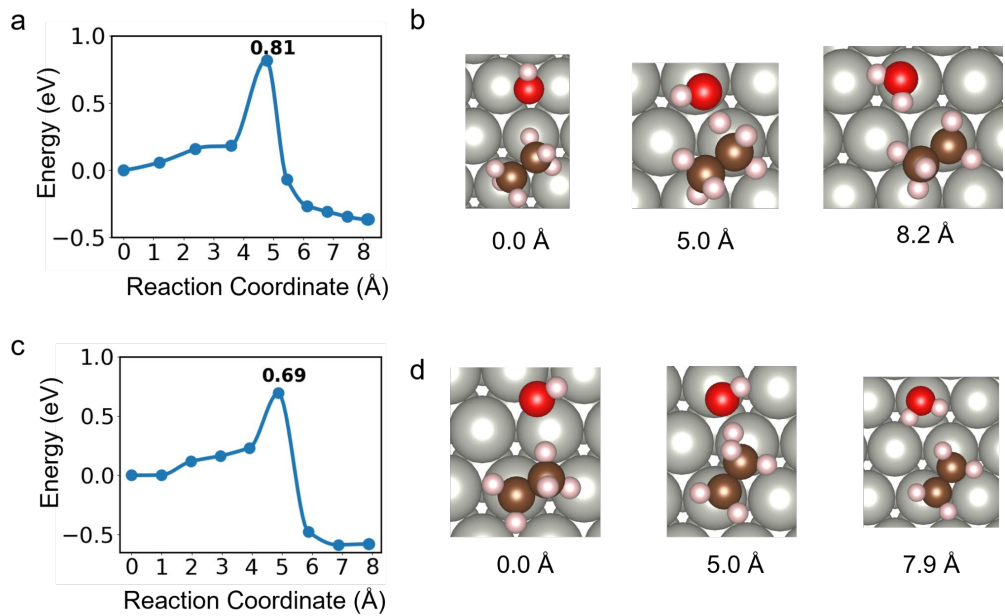

Supplementary Figure 13. Reaction pathways and visualization of surface-bound reaction of $C_2H_6 + OH \rightarrow C_2H_5 + H_2O$ (a,b) and $C_2H_5 + OH \rightarrow C_2H_4 + H_2O$ (c,d). Panels a and c (left column) show the plane-wave DFT (PW-PBE-D3BJ) energies along the CI-NEB MEP as a function of the reaction coordinate (in Å). The corresponding panels b and d (right column) display the structural models of the initial, transition, and final states along the MEP with the corresponding reaction coordinate. Gray, brown, tan, and red spheres represent Pd, C, H, and O atoms, respectively.

|  | Pd-O (CN * $S_0^2$) | Pd-O bond length(Å) | Pd-O Debye-Waller factor | Pd–Au (CN * $S_0^2$) | Pd-Au bond length (Å) | Pd-Au Debye-Waller factor |
|---|---|---|---|---|---|---|
| Pd/TiO$_2$ | 2.3 | 2.044 | 0.00171 |  |  |  |
| AuPd$_{0.01}$ |  |  |  | 8.0 | 2.825 | 0.00699 |
| AuPd$_{0.05}$ |  |  |  | 7.4 | 2.830 | 0.00924 |
| AuPd$_{0.1}$ | 1.3 | 2.003 | 0.00510 | 4.4 | 2.814 | 0.00984 |
| AuPd$_{0.2}$ | 1.7 | 2.043 | 0.00569 | 2.4 | 2.839 | 0.00756 |
| AuPd$_{0.3}$ | 1.6 | 2.032 | 0.00321 | 1.4 | 2.822 | 0.00631 |
| Pd foil |  |  |  | 12.0(Pd-Pd) | 2.740(Pd-Pd) | 0.00484(Pd-Pd) |

Supplementary Table 1. Bond length and Coordination number (CN) of the Pd local environment fitted from the extended X-ray absorption Fine Structure (EXAFS) spectrum. Amplitude reduction factor ($S_0^2$) is fixed at 0.742 using the EXAFS fitting results from the Pd foil collected simultaneously as a reference with a fixed first shell coordination number of Pd-Pd. The fitting errors for CN are around ±20%, for the bond length are within ±0.5%

| Wavenumber (cm⁻¹) | Vibration Mode | Species Assignment | Reference Source |
|---|---|---|---|
| ~3733, ~3697 | $\nu$(O-H) stretch | Terminal free OH* on $TiO_2$ surface | 1,2 |
| ~3670 | $\nu$(O-H) stretch (bridging) | Bridging OH* on $TiO_2$ surface | 2,3 |
| ~3610 | Molecular adsorbed OH* | Surface non-bridging OH* | 2,3 |
| 3500–3200 (Broad) | $\nu$(O-H) stretch | Adsorbed OH* with hydrogen bonds | 1,2 |
| ~1780–1720 | $\nu$(C=O) stretch | Two different chemical environment C=O | 4–6 |
| 1640 | $\delta$(H-O-H) scissoring/bending | Adsorbate $H_2O$ | 4–6 |
| ~1600-1470 | $\nu$(C=C) stretch | Unsaturated $CH_x$* (Aromatic C=C) | 4–6 |
| ~1596 | $\nu$(C-O) stretch | Bidentate Carbonate ($CO_3^{2-}$) | 7 |
| ~1595 | $\nu_{as}$(COO) stretch | Acetate ($CH_3COO^-$) | 8 |
| ~1472 | $\nu_{as}$(COO) or d(C-H) | Formate ($HCOO^-$) | 7 |
| ~1450 | $\delta_{as}$(CH3) | Surface adsorbed $CH_3$* | 5 |
| ~1440-1400 | $\delta$(O-H) in-plane bending | Carboxylic Acid (COOH*) | 4–6 |

Supplementary Table 2. Assignment of vibrational modes of the adsorbate species.

**Reference**

1. Gholami, R., Alyani, M. & Smith, K. Deactivation of Pd catalysts by water during low temperature methane oxidation relevant to natural gas vehicle converters. *Catalysts* **5**, 561–594 (2015).

2. *IR Spectroscopy of Surface Water and Hydroxyl Species on Nanocrystalline TiO2 Films*.

3. Yu, W., Zhao, L., Chen, F., Zhang, H. & Guo, L.-H. Surface Bridge Hydroxyl-Mediated Promotion of Reactive Oxygen Species in Different Particle Size TiO Suspensions. *J Phys Chem Lett* **10**, 3024–3028 (2019).

4. Socrates, G. *Infrared and Raman Characteristic Group Frequencies: Tables and Charts*. (John Wiley & Sons, 2004).

5. Vyvyan, J. R., Pavia, D. L., Lampman, G. M. & Kriz, G. S. *Introduction to Spectroscopy*. (Brooks/Cole, Florence, KY, 2013).

6. Nakamoto, K. *Infrared and Raman Spectra of Inorganic and Coordination Compounds, Part B*. (Wiley-Interscience, Newy York, 2009).

7. Li, Z., Xu, G. & Hoflund, G. B. In situ IR studies on the mechanism of methane oxidation over Pd/Al2O3 and Pd/Co3O4 catalysts. *Fuel Process. Technol.* **84**, 1–11 (2003).

8. Jiménez, J. D. *et al.* Identification of highly selective surface pathways for methane dry reforming using mechanochemical synthesis of Pd-CeO2. *ACS Catal.* **12**, 12809–12822 (2022).